\documentclass[fleqn,usenatbib]{mnras}

\usepackage[utf8]{inputenc}

\usepackage{acronym}
\usepackage{ae,aecompl}
\usepackage{booktabs}
\usepackage{dcolumn}
\usepackage{graphicx}
\usepackage{amsmath,amsfonts,amssymb}
\usepackage{mathrsfs}
\usepackage[normalem]{ulem}
\usepackage{epstopdf}

\usepackage[T1]{fontenc}

\let\oldhref\href
\renewcommand{\href}[2]{\oldhref{#1}{\hbox{#2}}}

\title[A systematic study of PNS convection in 3D CCSN]{A systematic study of proto-neutron star convection in three-dimensional core-collapse supernova simulations}

\author[H. Nagakura et al.]{
Hiroki Nagakura$^{1}$\thanks{E-mail: hirokin@astro.princeton.edu},
Adam Burrows$^{1}$,
David Radice$^{2,3}$,
David Vartanyan$^{4}$
\\
$^{1}$Department of Astrophysical Sciences, Princeton University, 4 Ivy Lane, Princeton, NJ 08544, USA\\
$^{2}$Department of Physics, The Pennsylvania State University, University Park, PA 16802, USA\\
$^{3}$Department of Astronomy \& Astrophysics, The Pennsylvania State University, University Park, PA 16802, USA\\
$^{4}$Astronomy Department and Theoretical Astrophysics Center, University of California, Berkeley, CA 94720, USA
}

\date{Accepted XXX. Received YYY; in original form ZZZ}

\pubyear{2019}

\begin{document}
\label{firstpage}
\pagerange{\pageref{firstpage}--\pageref{lastpage}}
\maketitle

\begin{abstract}
This paper presents the first systematic study of proto-neutron star (PNS) convection 
in three dimensions (3D) based on our latest numerical F{\sc{ornax}} models of core-collapse 
supernova (CCSN). We confirm that PNS convection commonly occurs, and then quantify the 
basic physical characteristics of the convection. By virtue of the large number of 
long-term models, the diversity of PNS convective behavior emerges. We find that the 
vigor of PNS convection is not correlated with CCSN dynamics at large radii, but rather 
with the mass of PNS $-$ heavier masses are associated with stronger PNS convection. We 
find that PNS convection boosts the luminosities of $\nu_{\mu}$, $\nu_{\tau}$, $\bar{\nu}_{\mu}$, 
and $\bar{\nu}_{\tau}$ neutrinos, while the impact on other species is complex due to a 
competition of factors. Finally, we assess the consequent impact on CCSN dynamics
and the potential for PNS convection to generate pulsar magnetic fields.
\end{abstract}


\section{Introduction}\label{sec:intro} 
After years of sustained effort, the core-collapse supernova (CCSN) 
community has recently fielded three-dimensional (3D) models of CCSN with detailed microphysical 
inputs and neutrino transport. Progress has been made through increased 
computational resources and improved numerical algorithms, both of which are still evolving. 
An important consequence of this progress is that numerous 3D models with varying degrees of sophistication 
have witnessed shock revival via the neutrino-driven explosion mechanism
\citep{2014ApJ...786...83T,2015ApJ...807L..31L,2016ApJ...831...98R,2018ApJ...865...61G,2018ApJ...855L...3O,2019arXiv190602009S,2019ApJ...873...45G,2019MNRAS.484.3307M,2019MNRAS.482..351V,2019MNRAS.485.3153B,2019MNRAS.tmp.2343N,2019arXiv190904152B} $-$ 
the only spherically-symmetric (1D) models that explode are those for very low-mass progenitors 
\citep{2006A&A...450..345K,2017ApJ...850...43R}.  Among other things, 3D explosion models allow 
us to quantify the expected observable neutrino and gravitational wave signals, and the detection
of those signals will provide vital clues to the explosion mechanism.

Recently, a large number of 3D simulations have become available and they reveal a rich 
diversity of explosion dynamics and observational signatures 
\citep{2019arXiv190904152B,2019MNRAS.489.2227V,2019ApJ...876L...9R}. Low-mass 
progenitors tend to generate less energetic explosions (at least on average), whereas high-mass 
progenitors seem to explode more energetically, if the shock is revived and 
the binding energy of the outer envelope is not too large. However, there are suggestions 
that some progenitors are less prone to explosion
(\citet{Pan:2017tpk,2018ApJ...855L...3O,2018MNRAS.477L..80K,2019arXiv191012971W}). 
Once it achieves the maximum mass limit, the proto-neutron star (PNS) formed in that 
context will collapse to a black hole. Whatever happens, the internal dynamical behavior 
of the event and aspects of the nuclear equation of state (EOS) will be 
imprinted on its neutrino and gravitational wave signals \citep{2006PhRvL..97i1101S,2016MNRAS.461.3296N,2018ApJ...861...10M,2019PhRvL.123e1102T,2019ApJ...876L...9R,2019arXiv190911816S}. 
Be that as it may, the large number of 3D CCSN simulations with 
state-of-the art physics that are emerging in the published literature are opening
a new window into the explodability and observational signals of massive stars.

The physics of inner PNS convection has been studied for many years \citep{2006ApJ...645..534D} 
and is one of the key elements of CCSN theory. 
Such convection appears in the inner regions ($10 \lesssim r \lesssim 25$ km) of the PNS
and persists whether the shock wave is revived or not $-$ it is a generic feature in all stellar 
collapse scenarios.  It is driven predominantly by negative lepton gradients that
are a consequence of net and persistent electron-neutrino loss and is sustained until the PNS 
has almost fully deleptonized and cooled many seconds after bounce \citep{1986ApJ...307..178B,2012PhRvL.108f1103R}. 
This characteristic timescale is much longer that the dynamical timescale of the 
PNS ($\sim$1 ms), and the PNS evolves quasi-hydrostatically.  Though neutrino-driven
convection behind the shock has in the past been studied as a possible agency and site 
for the dynamo origin of pulsar magnetism \citep{1993ApJ...408..194T,2006A&A...451.1049B}, 
PNS convection in the inner region is also a possible context for dynamo action and 
magnetic field generation.  This possibility, that pulsar B-fields are sourced in
the inner proto-neutron star, is one motivation for the closer
look at the physics and phenomena that accompany PNS convection represented by this paper.

Moreover, the role of PNS convection on CCSN dynamics is still an open question. 
In earlier studies, it was suggested that it played a direct role in
the explosion mechanism \citep{1979MNRAS.188..305E,1979ApJ...234L.183B,1980ApJ...238L.139L}. PNS
convection was supposed to generate large-scale overturn, significantly enhancing neutrino fluxes 
and dramatically increasing neutrino heating in the outer gain region. However, as 
pointed out by \citet{1981ApJ...246..955L} and \citet{1988PhR...163...51B}, the positive entropy 
gradient, which coexists with the negative lepton gradient in the envelope of 
the PNS, stabilizes convection there and suppresses large-scale overturn.  
Furthermore, \citet{1995ApJ...450..830B} observed neither large-scale overturn, nor enhancement of neutrino 
luminosities, in their axisymmetric simulations. On the other hand, \citet{1996ApJ...473L.111K} 
found that PNS convection increases both neutrino luminosities and the average neutrino energies 
by factors of $\sim$2 and $\sim$10\%, respectively, compared with spherically-symmetric 
models. It should be noted, however, that those numerical simulations employed 
a crude gray approximation to neutrino transport and this is likely a crucial deficit. 
Indeed, \citet{1998ApJ...493..848M} suggested that the growth of PNS convection is 
delayed when a better treatment of neutrino transport is employed, suggesting that the 
treatment of neutrino transport may change the characteristics of PNS convection.

Detailed multi-group studies of PNS convection in axisymmetry were conducted by 
\citet{2006ApJ...645..534D} and \citet{2006A&A...457..281B}. The former showed that the 
effect of PNS convection on neutrino emission depends upon neutrino 
species; $\bar{\nu}_e$ (electron-type anti-neutrino) and $\nu_x$ (heavy-leptonic neutrino) 
luminosities were roughly $\gtrsim 15 \%$ and $\gtrsim 30 \%$ higher, respectively, 
than those in spherically-symmetric models and $\nu_e$ (electron-type neutrino) 
emissions were less sensitive to PNS convection. However, the 
latter study suggested that the neutrino luminosities of all species are enhanced 
by PNS convection. In addition, it has recently been suggested that PNS convection could 
influence shock revival by the generation of outgoing hydrodynamic waves \citep{2019arXiv191007599G},
contributing perhaps $\sim$10\% to the power deposition behind the shock. It 
should be noted, however, that sound waves generated by infalling matter plumes that impinge 
upon the PNS, could be equally in play, though both processes are likely subdominant.  
To be sure, these mechanisms remain to be adequately quantified.

In this paper, we present results of the first systematic study, across a broad range of 
massive-star progenitors and supernova models, of PNS convection using the results of state-of-the 
art 3D CCSN simulations. A few previous studies addressed PNS convection in 3D 
\citep{2019ApJ...881...36G,2019arXiv190106235W,2019MNRAS.487.1178P}, but they were limited to 
a few models. Moreover, those studies generally focused on the relation between PNS convection and
the lepton-emission self-sustained asymmetry (LESA) phenomenon 
\citep{2014ApJ...792...96T,2018ApJ...865...81O,2019MNRAS.489.2227V}) 
or gravitational radiation \citep{2019MNRAS.487.1178P}. In this study, we integrate the 
results of sixteen 3D CCSN simulations computed by F{\sc{ornax}} \citep{2019arXiv190904152B}. 
By using a large number of models, we address the following fundamental questions 
in PNS convection: what is the progenitor dependence?; are there any qualitative differences between 
exploding and non-exploding models?; and what are the dynamical features of PNS convection? 
In addressing these issues, we quantify the basic characteristics and consequences of PNS convection.
In Sec.~\ref{sec:methodandmodel}, we briefly summarize our 3D 
numerical CCSN models. We present the results of our analyses in 
Sec.~\ref{sec:results}. This is the core of the paper. Finally, we summarize 
and discuss our results in Sec.~\ref{sec:sumanddiscuss}.
We stress that PNS convection is not the ``neutron-finger" convection posited by
J. Wilson \citep{1985ApJ...295...14B,1988PhR...163...63W}.  Such doubly-diffusive instabilities
have been shown not to arise in the supernova context \citep{1996ApJ...458L..71B,2006ApJ...645..534D}.
Rather, this is classic Ledoux convection, but in the exotic environment of the PNS core.

\begin{figure}
  \begin{minipage}{1.0\hsize}
    \includegraphics[width=\columnwidth]{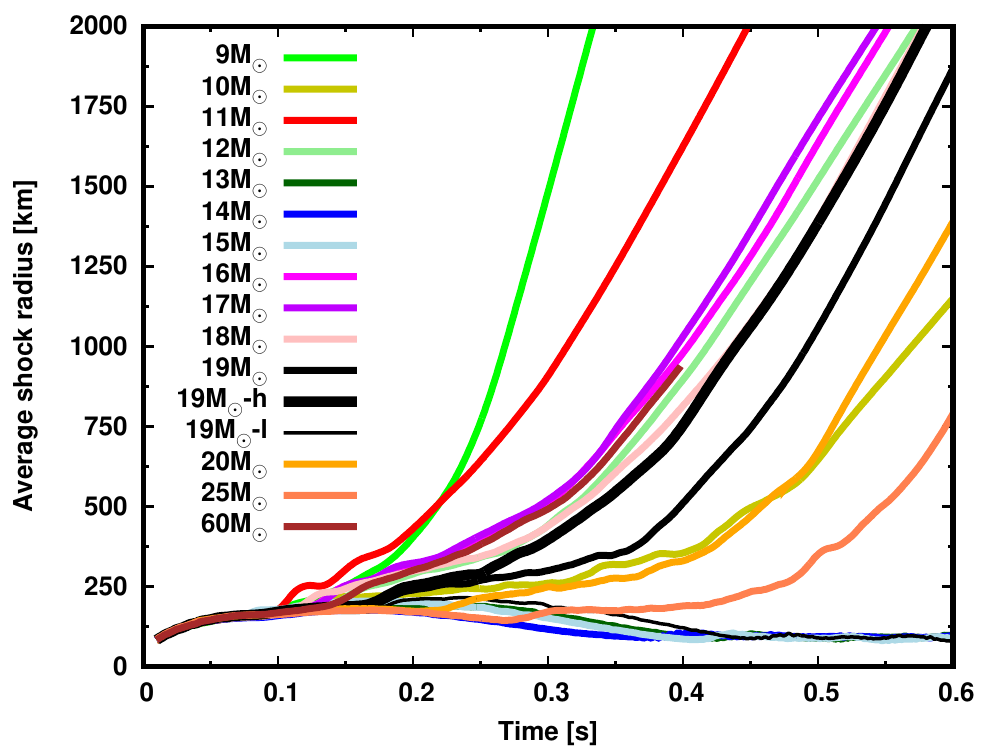}
    \caption{The time evolution of the mean shock radius. Color denotes the ZAMS mass of the progenitor. 
For the $19-M_{\sun}$ model (black line), the results of high and low resolution models are displayed 
as thicker and thinner lines, respectively, than the line denoting the standard resolution model.}
    \label{graph_timetrajectories_shockradii_PNS}
  \end{minipage}
\end{figure}

\begin{figure}
  \begin{minipage}{1.0\hsize}
    \includegraphics[width=\columnwidth]{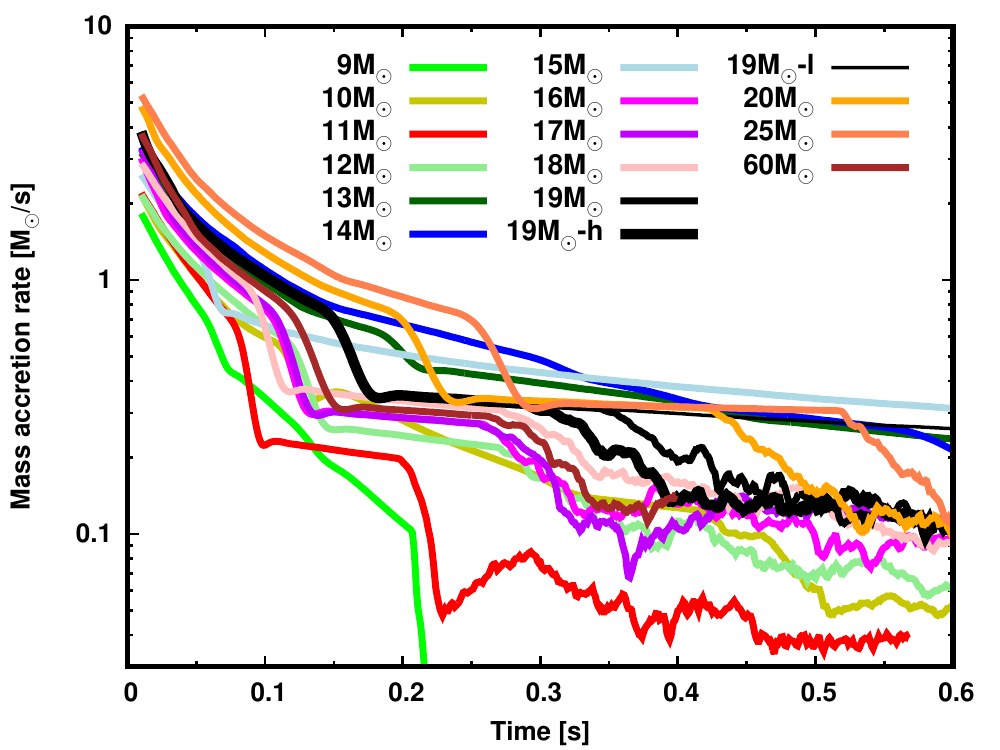}
    \caption{The time evolution of the mass accretion rate measured at 500 km for all 3D models 
presented in this paper.}
    \label{graph_Massaccretion}
  \end{minipage}
\end{figure}

\begin{figure}
  \begin{minipage}{1.0\hsize}
    \includegraphics[width=\columnwidth]{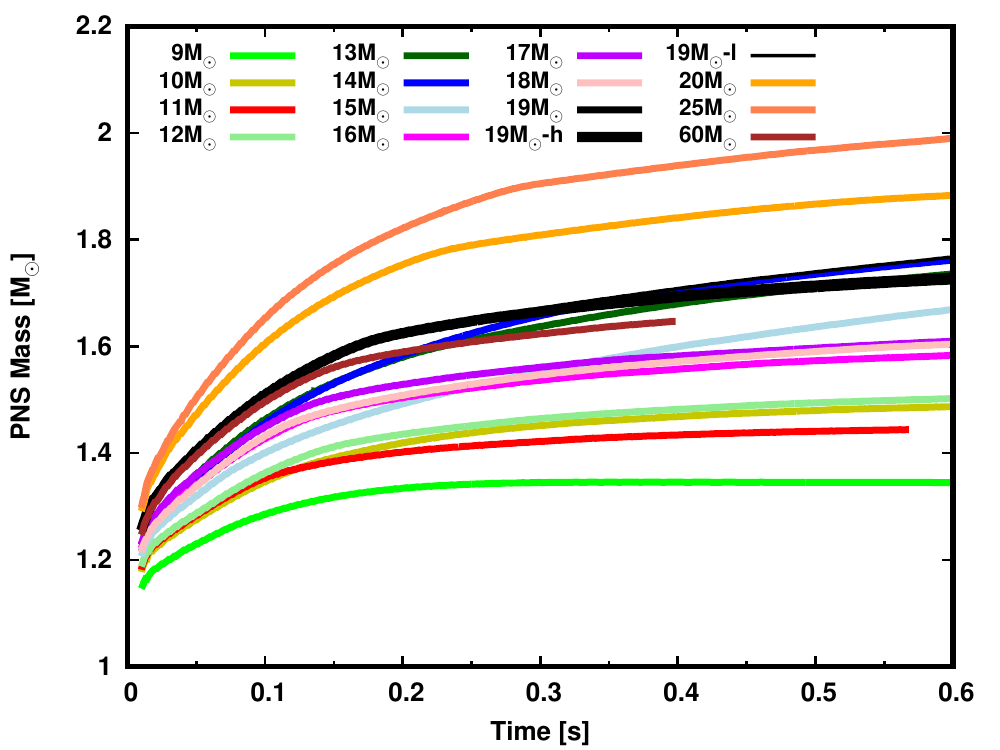}
    \caption{The time evolution of the mass of the PNS, which is defined as the enclosed baryon mass up 
to a radius at which the angle-averaged mass density ($\rho_{\rm ave}$) is $10^{11} {\rm g/cm^3}$. 
As discussed in the body of the text, the mass of the PNS has a clear correlation 
with the vigor of PNS convection (see, e.g., Fig.~\ref{graph_tevo_PNSconv_kineticEne}).}
    \label{graph_PNSmass}
  \end{minipage}
\end{figure}

\begin{figure}
  \begin{minipage}{1.0\hsize}
    \includegraphics[width=\columnwidth]{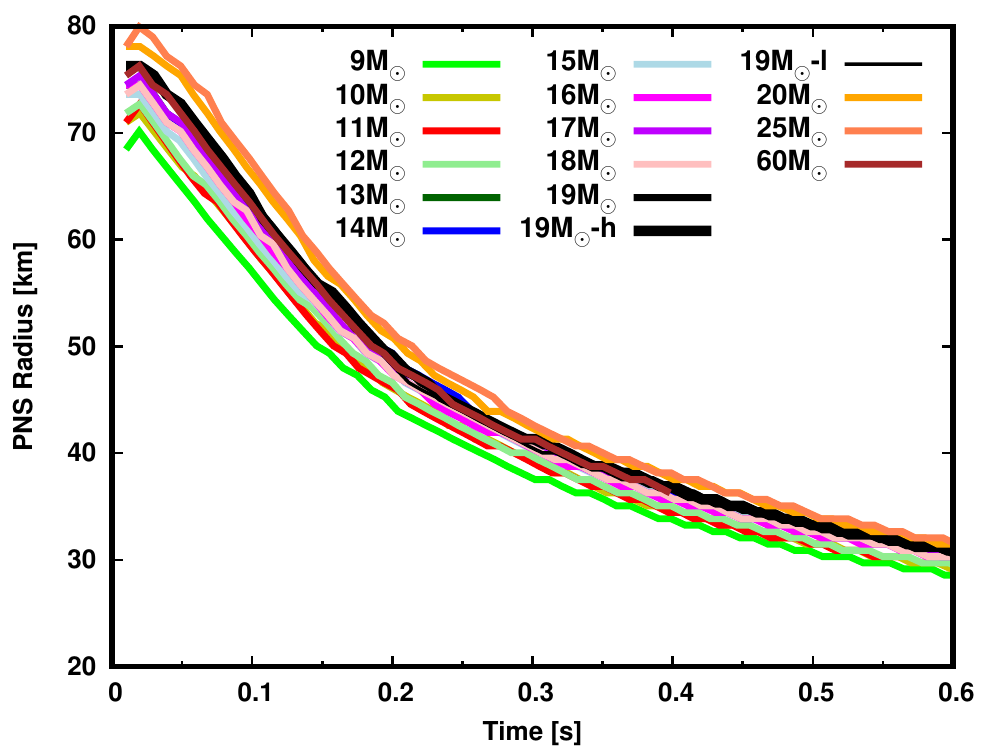}
    \caption{The time evolution of the PNS radius, defined at the radius where $\rho=10^{11} {\rm g/cm^3}$ on average.}
    \label{graph_PNSradius}
  \end{minipage}
\end{figure}

\section{Method and Model}\label{sec:methodandmodel}

All 3D CCSN simulations presented in this paper were carried out using our 3D neutrino-radiation 
hydrodynamic code, F{\sc{ornax}}. The details of the methodology and basic equations 
and code tests can be found in a series of our previous papers 
\citep{2016ApJ...831...81S,2017ApJ...850...43R,2018MNRAS.477.3091V,2019ApJS..241....7S}. 
We also direct the reader to \citet{2019arXiv190904152B} for a discussion of the input 
physics and numerical setups employed. Here, we highlight the results for fourteen solar-metallicity, 
non-rotating CCSN progenitors with zero-age-main sequence (ZAMS) masses from 
$9$ to $20\, M_{\sun}$, taken from \citet{2016ApJ...821...38S}. We also include a $60\, M_{\sun}$ progenitor from the same 
reference and the $25\, M_{\sun}$ model from \citet{2018ApJ...860...93S}. We start all simulations in spherically 
symmetry and map them to 3D at $10$ms after core bounce, adding very modest velocity 
perturbations (with maximum speed $100\, {\rm km/s}$), following the prescription detailed in 
\citet{2015MNRAS.448.2141M}. We adopt spherical coordinates out to $20,000$ km from the stellar center by deploying 
$678 (r) \times 128 (\theta) \times 256 (\phi)$ grid points as our standard resolution.  We also consider
high- and low-resolution simulations of the $19\, M_{\sun}$ progenitor, in which the angular 
resolution in the $\theta$ and $\phi$ directions is higher and lower by a factor of two 
than the default \citep{2019MNRAS.tmp.2343N}. Therefore, in total, we consider sixteen CCSN models in 
this study. We use a common spatial grid for the hydrodynamics and the neutrino transport. For 
the latter, we employ $12$ energy groups and three neutrino species are distinguished: $\nu_e$, 
$\bar{\nu}_e$, and $\nu_x$. Note that we assume that all heavy leptonic neutrinos and their anti-partners 
have the same distribution functions.

The detailed analyses of the explodability and of the observable 
characteristics imprinted in gravitational waves and neutrinos of these models can 
be found in \citet{2019arXiv190904152B}, \citet{2019ApJ...876L...9R} and \citet{2019MNRAS.489.2227V}.
As shown in Fig.~\ref{graph_timetrajectories_shockradii_PNS}, the majority of our models explode, 
but our current 3D models with $13-, 14-$ and $15\, M_{\sun}$ progenitors and our low-resolution model 
of the $19\, M_{\sun}$ progenitor do not by the end of those simulations. We note 
that the $19\, M_{\sun}$ models with standard and higher angular resolutions experienced shock 
revival, the higher-resolution model experiencing more vigorous shock expansion than that 
of the standard resolution (see the thick black line in Fig.~\ref{graph_timetrajectories_shockradii_PNS}). 
Also, as we show in \S\ref{sec:results}, the different $19\, M_{\sun}$ models provide insight into the potential
effect of shock revival on PNS convection.

In support of our general description of PNS convection, we provide the time evolution of the 
mass-accretion rates (Fig.~\ref{graph_Massaccretion}), PNS masses 
(Fig.~\ref{graph_PNSmass}), and PNS radii (Fig.~\ref{graph_PNSradius}).  These plots 
support our description of PNS convection and its systematics with progenitor. One of 
remarkable trend is that the mass accretion rate of the $25\, M_{\sun}$ and 
$9\, M_{\sun}$ models is the highest and lowest, respectively, among all models, the PNS mass 
recapitulates this trend, and that the PNS radius trends similarly, though with 
a smaller spread. We show in the next section that the vigor of PNS 
convection correlates closely with PNS mass, whereas the dependence of PNS convection 
overall on explodability, mass accretion rate, and PNS radius is weaker.

\section{Results}\label{sec:results}
In this section, we present our central results. In \S\ref{subsec:Prodepe}, after describing the basic 
characteristics of PNS convection, we delve into the diversity with progenitor. In \S\ref{subsec:turbulence}, 
we analyze the characteristics of the turbulence in PNS convection. Finally, in 
\S\ref{subsec:neutrinoemis} we discuss the impact of PNS convection on neutrino emissions.

\subsection{Basic characteristics of PNS convection}\label{subsec:basicchara}

\begin{figure*}
  \rotatebox{90}{
    \begin{minipage}{1.2\hsize}
	\includegraphics[width=\columnwidth]{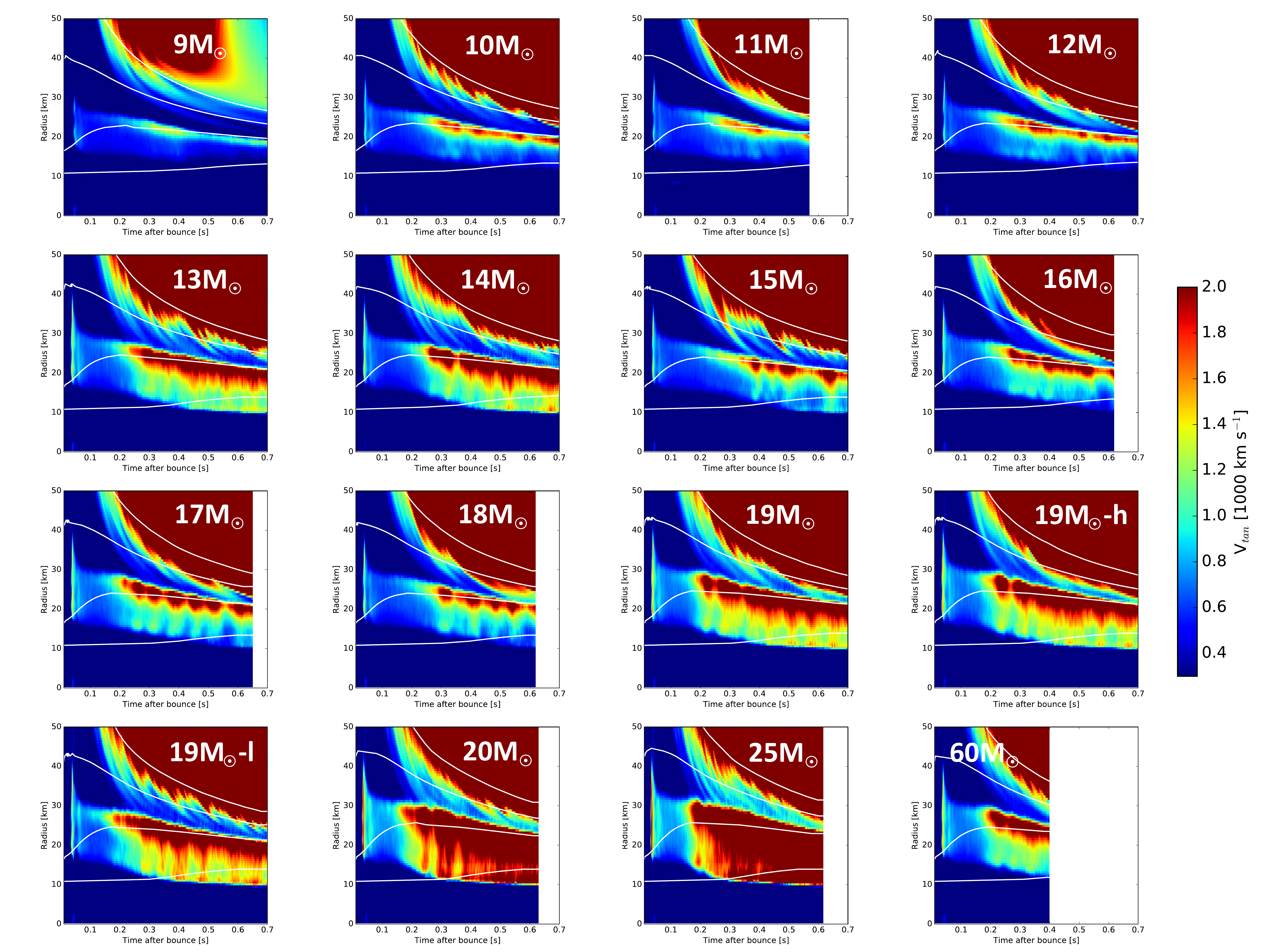}
    \caption{Radius-time diagrams of the angle-averaged speed of non-radial fluid motions ($V_{\rm tan}$) 
in the inner 50 km through $0.7$ s. Models terminated before $0.7$s contain blank white 
space from the time at the end of the simulation to $0.7$s. In each panel, there are two regions 
with high $V_{\rm tan}$ (brown), but only the bottom one corresponds to the region experiencing 
PNS convection. The other regions are the base of the outer convective zone driven by neutrino heating 
from below. On each colormap as white lines, we also display angled-average isodensity radii, 
delineating $10^{11}, 10^{12}, 10^{13}$ and $10^{14} {\rm g/cm^3}$, respectively.}
    \label{LateralVmap}
  \end{minipage}}
\end{figure*}

\begin{figure*}
  \rotatebox{0}{
    \begin{minipage}{0.9\hsize}
	\includegraphics[width=\columnwidth]{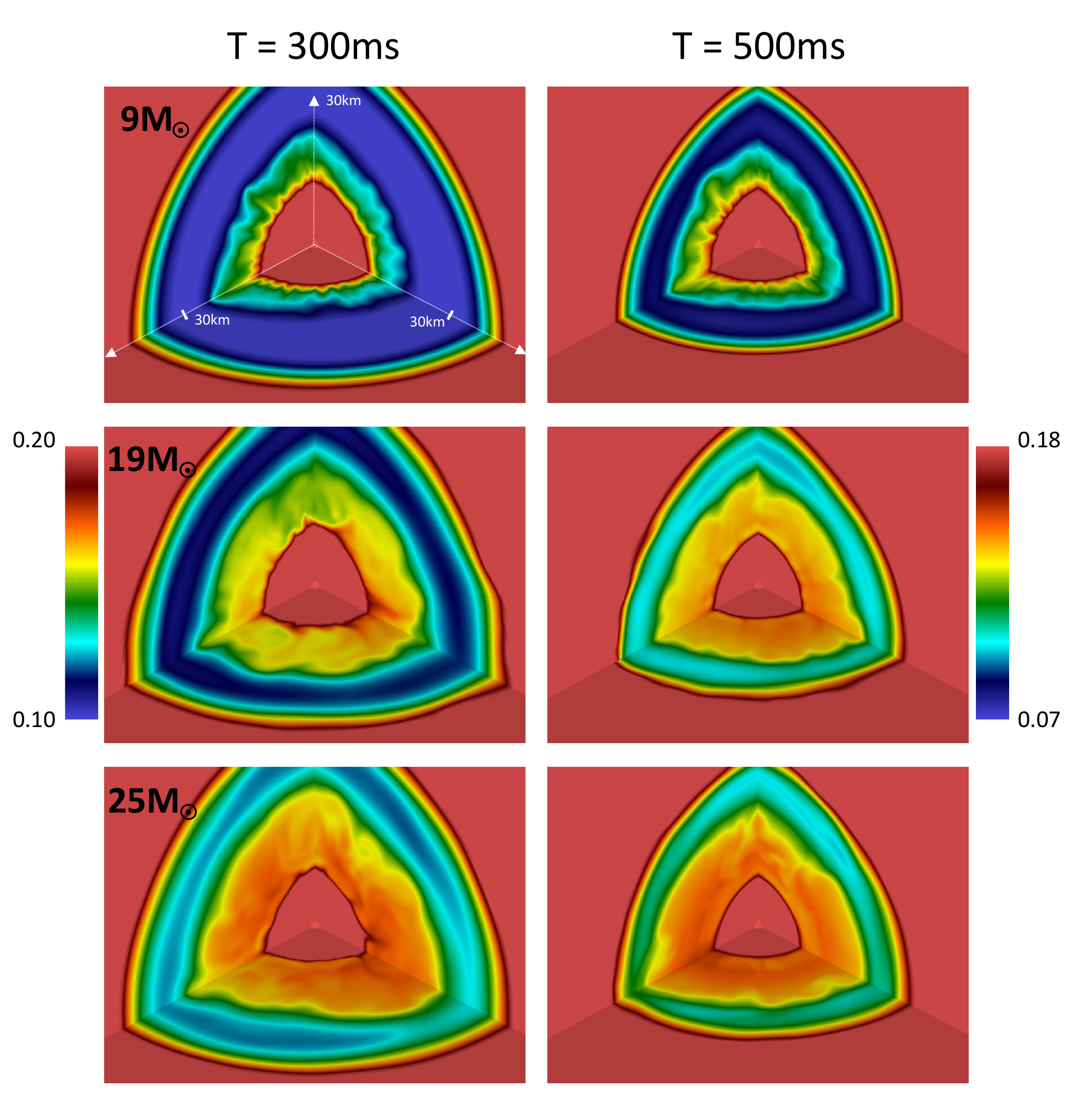}
    \caption{3D color maps of the $Y_e$ distributions in the vicinity of PNS. The left and right panels correspond 
to post-bounce times of $t=300$ms and $t=500$ms, respectively. From top to bottom, the results for 
the $9-, 19-$ and $25-M_{\sun}$ models are displayed.}
    \label{graph_3Dvisualization}
  \end{minipage}}
\end{figure*}

\begin{figure*}
  \begin{minipage}{0.9\hsize}
    \includegraphics[width=\columnwidth]{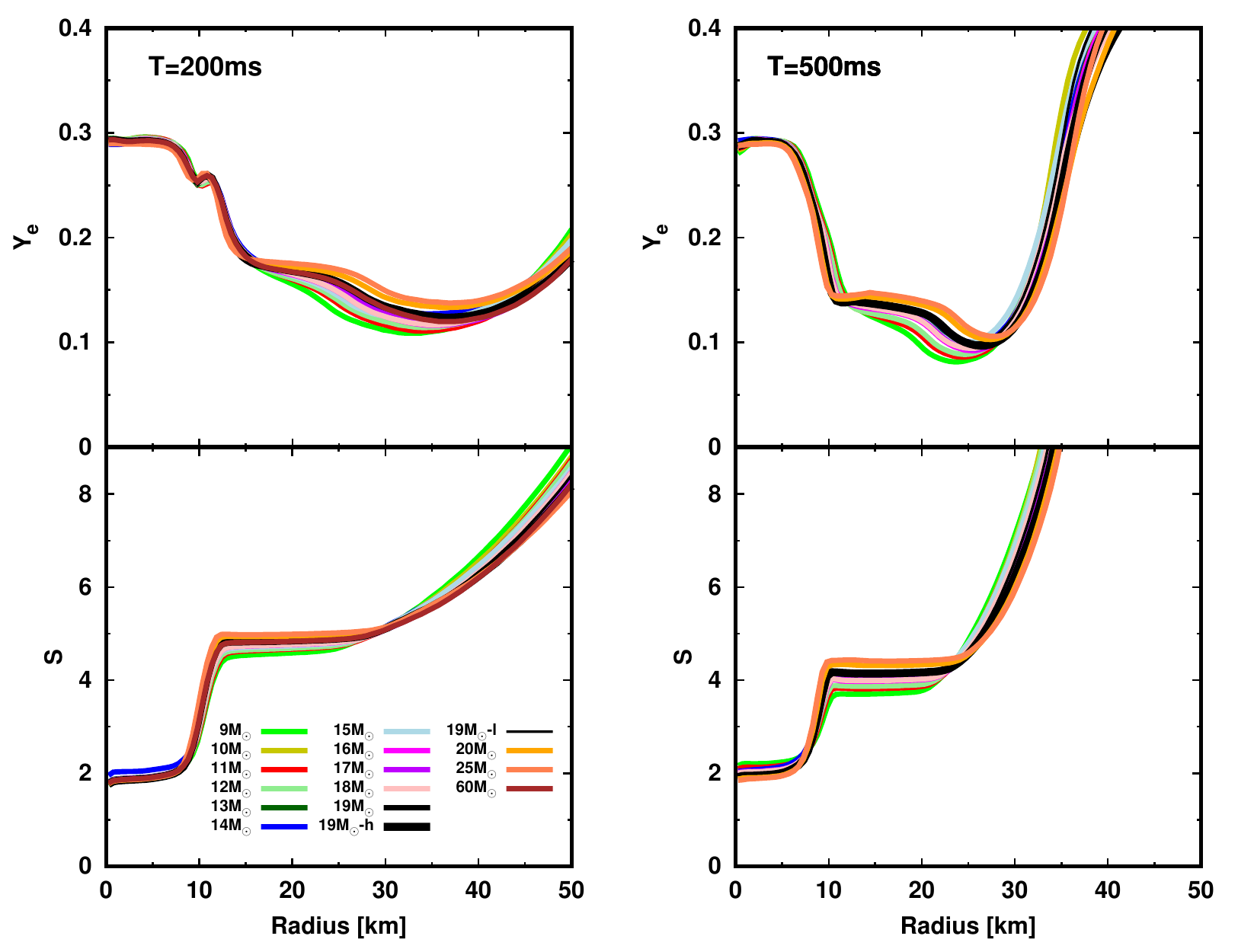}
    \caption{Radial profiles of angle-averaged electron fraction ($Y_e$) on top and entropy per baryon 
(S) on bottom panels at $200$ ms (left panel) and $500$ ms (right panel) after bounce, respectively. On the 
right panel, we do not include the $60\, M_{\sun}$ model.}
    \label{graph_Ye_S_profile_asR_t200ms500ms}
  \end{minipage}
\end{figure*}


\begin{figure}
  \begin{minipage}{1.0\hsize}
    \includegraphics[width=\columnwidth]{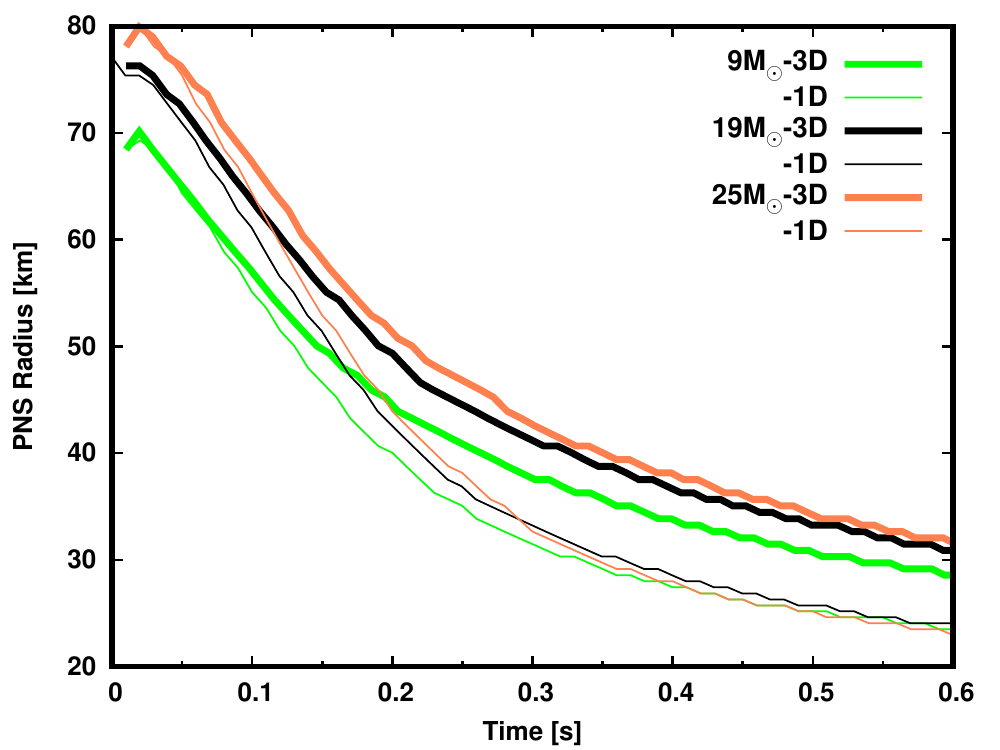}
    \caption{Comparison of the time evolution of the PNS radius between 3D (thick lines) 
and 1D (thin lines) models for the three representative models.}
    \label{graph_PNSradius_1Dvs3D}
  \end{minipage}
\end{figure}

\begin{figure}
  \begin{minipage}{1.0\hsize}
    \includegraphics[width=\columnwidth]{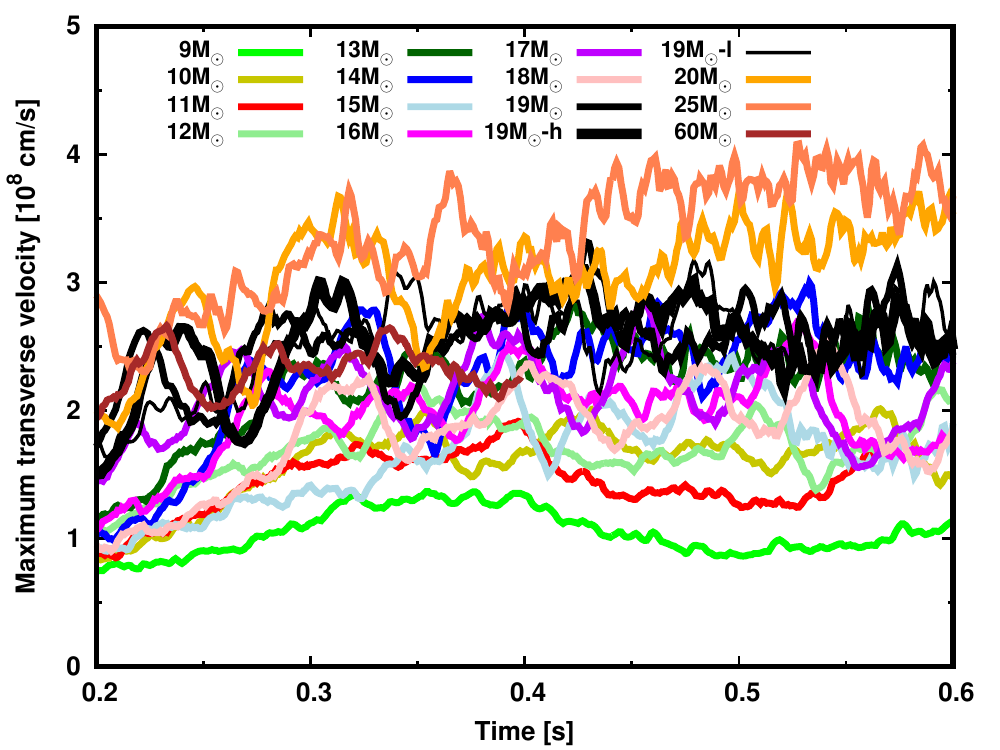}
    \caption{Time evolution of the maximum tangential velocity in the region of PNS convection.}
    \label{graph_tevo_vtanmax}
  \end{minipage}
\end{figure}

Figure~\ref{LateralVmap} displays radial distributions of the angle-averaged speed of non-radial, transverse 
fluid motions ($V_{\rm tan}$) for all models as a function of time in the region bounded 
by $10 \lesssim r \lesssim 25$ km in which PNS convection is manifest
and for which $V_{\rm tan} \gtrsim 10^{8}\, {\rm km/s}$. PNS convection 
commences $\lesssim 50$ ms after bounce and is sustained during the rest 
of the post-bounce phase. The Mach number associated with the convective motion is an order of $\sim 10^{-2}$, 
which is of the same order as that found in previous 2D simulations \citep{2006ApJ...645..534D,2006A&A...457..281B}. 
Commonly, the convective layer moves inward with time in accordance with PNS contraction. 
We observe that the convective layer resides between the two isodensity surfaces 
$10^{13}$ and $10^{14}\, {\rm g/cm^3}$, except during the very early post-bounce phase. The spatial width 
of this convective layer is $\sim\, 10$ km and is almost constant with time in our 
simulations\footnote{Note that the inner boundary of the convective layer sinks further inside the PNS core 
in the later phase and that the entire core will eventually become convective. This happens on the timescale 
of roughly a few seconds \citep{2012PhRvL.108f1103R}.}. The characteristic frequencies of 
PNS convection can be estimated as
\begin{eqnarray}
f \sim \frac{V_{tan} \ell}{\pi R} 
\sim 100 
\hspace{1mm} \left(\frac{V_{\rm tan}}{2 \times 10^8 {\rm cm/s}}\right)
\hspace{1mm} \left(\frac{\ell}{10}\right)
\hspace{1mm} \left(\frac{R}{20 {\rm km}}\right)^{-1}
\hspace{1mm} [{\rm Hz}]\, ,
\label{eq:freqeddyrela}
\end{eqnarray}
where $R$ and $\ell$ denote the mean radius of PNS convection and the spherical harmonic index, respectively. 
Note that $\ell$ is associated with the size of an eddy in the convective layer, which is induced 
in the non-linear phase by the turbulent cascade (see \S\ref{subsec:turbulence}). The small eddy motions 
are responsible for the higher frequencies.  The corresponding characteristic timescales range from $\sim$3 ms to $\sim$20 ms. 

We confirm that there is an annulus of weak non-radial motions in the region between 
the two isodensity surfaces at $10^{12}$ and $10^{13} {\rm g/cm^3}$, where the positive radial gradient in 
the entropy suppresses large-scale overturn (see also Fig.~\ref{graph_Ye_S_profile_asR_t200ms500ms}). 
The convectively stable layer is also observed in the $Y_e$ distributions, which are displayed 
in Fig.~\ref{graph_3Dvisualization}. The magnitude of the $Y_e$ fluctuations in the     
convective layer are $\sim$10\%, but such fluctuations are muted in the convectively 
stable region. 

Figure~\ref{graph_Ye_S_profile_asR_t200ms500ms} portrays the radial profiles of the angle-averaged electron fraction 
($Y_e$) and entropy per baryon (S) at two different time snapshots: $200$ ms and $500$ ms after bounce. 
Those distributions deep inside the PNS core ($r \lesssim 10$km) are approximately universal among the progenitors
\citep{1985ApJS...58..771B,2000PhR...333..121L,2003NuPhA.719..144L}. Such behavior is consistent with both  
1) the universal dynamics of the inner core during the collapse phase, in which both $Y_e$ and entropy 
distributions are self-regulated by electron-capture reactions \citep{2019ApJS..240...38N} and 2) 
to the fact that the timescale of neutrino diffusion inside the PNS core is of order seconds and the changes 
in $Y_e$ and entropy are nearly identical \citep{2019ApJS..240...38N}. 
The inner $Y_e$ and entropy distributions in the PNS convective layer are qualitatively 
similar from model to model. One of the salient (and expected) characteristics of these 
3D radial profiles is that they are flat in the convective layer. This is particularly true 
for the entropy profile, and is in sharp contrast with what obtains in spherically-symmetric models. 
Moreover, PNS convection dredges up the surface matter of the PNS core, and thereby 
smoothes the otherwise sharp bumps in the radial profile of $Y_e$.  It should be 
noted, however, that the $Y_e$ gradient on the outside of the PNS is still negative, and 
this continues to drive PNS convection. The negative $Y_e$ gradient is sustained by lepton-loss 
at the outer edge of the convective layer, in particular by lower-energy $\nu_e$s that have 
smaller interaction cross-sections.

We find that the convectively unstable region is always wider than the unstable regions in 1D. 
There are several reasons for this difference: 1) convective overshoot and the turbulent pressure, both of which 
enlarge the convective layer, are artificially suppressed in 1D; and 2) PNS expansion due to the 
accelerated deleptonization by PNS convection is suppressed in 1D \citep{2006A&A...457..281B}. 
In fact, the PNS radius is consistently larger in 3D than in 1D (Fig.~\ref{graph_PNSradius_1Dvs3D}). 
These characteristics also affect the progenitor dependence of PNS convection, which is discussed next.

\subsection{Progenitor dependence of PNS convection}\label{subsec:Prodepe}

\begin{figure}
  \begin{minipage}{1.0\hsize}
    \includegraphics[width=\columnwidth]{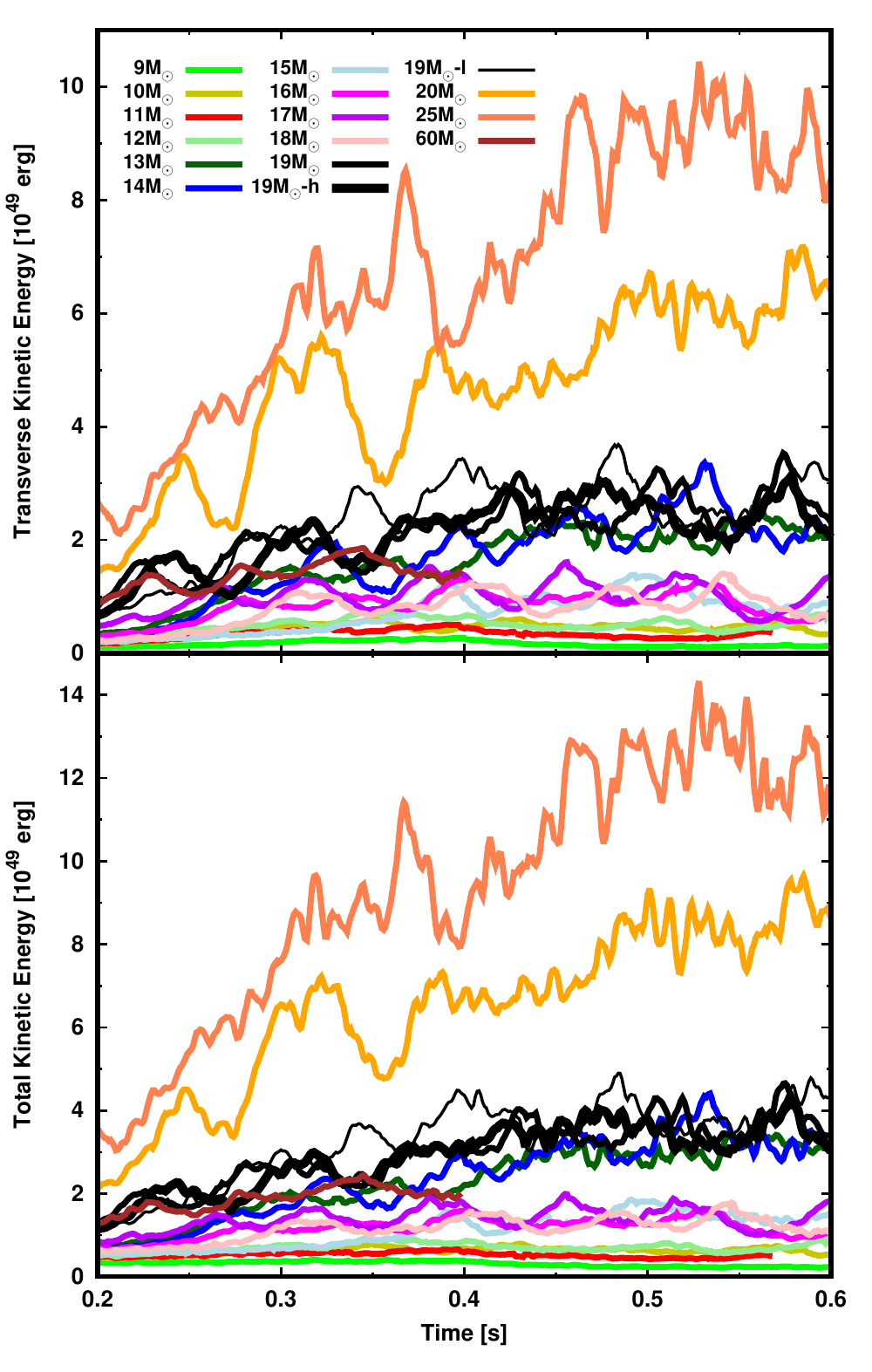}
    \caption{Time evolution of the transverse (upper) and total (bottom) kinetic energy 
of PNS convection for the 3D models of this study.}
    \label{graph_tevo_PNSconv_kineticEne}
  \end{minipage}
\end{figure}

\begin{figure}
  \begin{minipage}{1.0\hsize}
    \includegraphics[width=\columnwidth]{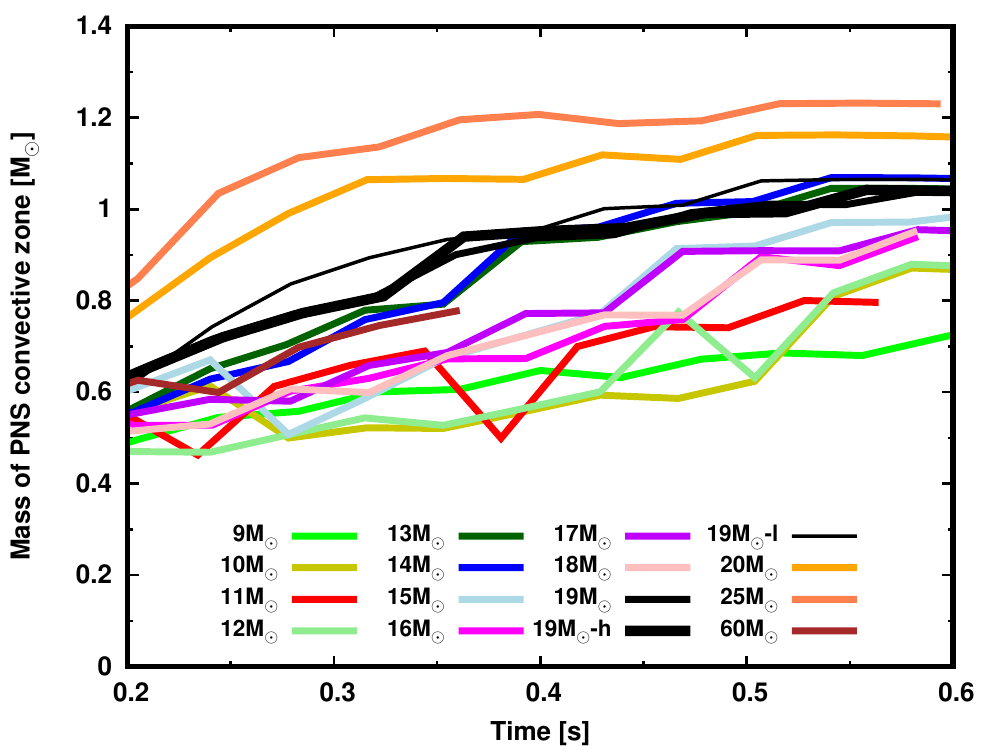}
    \caption{The time evolution of the baryon mass for all models in the region of PNS convection.
See text for a discussion.}
    \label{graph_tevo_PNSconv_Mass}
  \end{minipage}
\end{figure}

\begin{figure}
  \begin{minipage}{1.0\hsize}
    \includegraphics[width=\columnwidth]{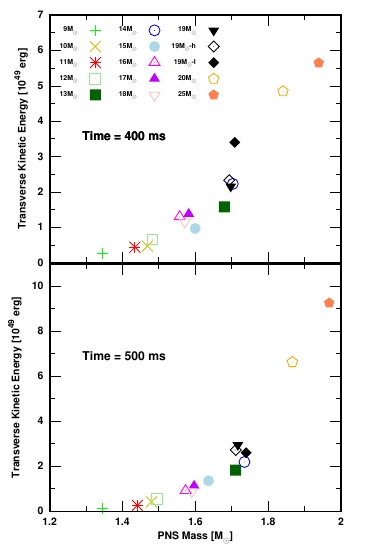}
    \caption{PNS mass vs. the transverse kinetic energy of PNS convection. The upper
and lower panels are for $T = 400$ and $500$ ms, respectively, after bounce.
The $60 M_{\sun}$ model is not included in these plots.}
    \label{graph_PNSmass_vs_kineticEne}
  \end{minipage}
\end{figure}


We now turn our attention to a more quantitative discussion of the diversity 
observed in PNS convection. Importantly, we find that low-mass progenitor models tend to have more quiet PNS 
convection. Indeed, the $9\, M_{\sun}$ model, which is the lowest ZAMS mass model among our model suite, 
has the weakest PNS convection (Figure~\ref{LateralVmap}). Such a correlation between the ZAMS mass and the
vigor of PNS convection is, however, not rigorously monotonic. For instance, the $10\, M_{\sun}$ model 
experiences a bit stronger convection than found in the $11\, M_{\sun}$ model. 

The $20\, M_{\sun}$ and $25\, M_{\sun}$ models, on the other hand, have wider and more vigorous 
convective layers than found in the other models.  Generally, models with wider convective zones experience
more vigorous PNS convection. This can be clearly seen by comparing the maximum tangential velocities 
in the region of PNS convection among models (Fig.~\ref{graph_tevo_vtanmax}). The 
$20\, M_{\sun}$ and $25\, M_{\sun}$ models have faster non-radial motions than the others. We reiterate 
that the $9\, M_{\sun}$ model has the lowest transverse convective speeds. 
The ratio of the maximum tangential velocity for the strongest and the weakest models
is roughly a factor of $4$, which indicates that the kinetic energy in the convective layer 
varies by more than an order of magnitude. As shown in Fig.~\ref{graph_tevo_PNSconv_kineticEne}, 
the total kinetic energy of PNS convection is roughly a few $10^{49} {\rm erg}$ but it can 
exceeds $10^{50} {\rm erg}$ for the most convective models.

We find that the vigor of PNS convection accounts in part for the difference among models in
the $Y_e$ distributions (Figures~\ref{graph_3Dvisualization}~and~\ref{graph_Ye_S_profile_asR_t200ms500ms}). 
Convection dredges up $Y_e$-rich matter inside the PNS core and carries the lepton-number outward
via large eddies. This process works more efficiently in the stronger convective context and results in higher
lepton number flux (see \S\ref{subsec:neutrinoemis}) and higher $Y_e$s in the outer PNS (due 
to the consequent increase in the absorption of $\nu_e$ by free neutrons).

Here, we briefly remark on the resolution dependence \citep{2019MNRAS.tmp.2343N}. Comparing 
three $19\, M_{\sun}$ models, we find that the general vigor of PNS convection is not very sensitive 
to resolution. More quantitatively, however, our low-resolution $19\, M_{\sun}$ model experiences slightly stronger 
convection than the higher-resolution model. It should be noted that the lowest resolution $19\, M_{\sun}$ 
model does not explode and continues to accrete matter. However, the other non-exploding models,
the $13-$, $14-$ and $15-M_{\sun}$ models, do not in any obvious way, despite their shared outcome, share 
any obvious features with the $19\, M_{\sun}$ models. 

Indeed, we find that the baryon mass of the PNS is the most important determinant of the vigor
of PNS convection. This trend can be seen by comparing two figures, 
Figs.~\ref{graph_PNSmass}~and~\ref{graph_tevo_PNSconv_kineticEne}, in which stronger convection 
is seen to be generated in models with a higher PNS mass\footnote{Note that the total mass of the PNS is here 
defined as the mass enclosed by an isodensity surface with $\rho = 10^{11} {\rm g/cm^3}$.}. 
This is more clearly shown in Fig.~\ref{graph_PNSmass_vs_kineticEne}, which compares for the various progenitors                                     
the PNS mass and the transverse turbulent kinetic energy in the convective region at $T=400$ and $500$ ms after bounce.
Consistently, models with vigorous convection tends to have higher baryon masses in the convective layer itself
(varying roughly by a factor of $\sim$ 2; see Fig.~\ref{graph_tevo_PNSconv_Mass}). Of course, the mass difference 
enhances the spread in the kinetic energy of PNS motions between strongly and 
weakly convective proto-neutron stars (Fig.~\ref{graph_tevo_PNSconv_kineticEne}).

The correlation between PNS mass and the vigor of PNS convection is one of the most 
important findings in this paper. It is in part a consequence of the deeper gravitational potential well
in models with a heavier PNS mass, which thereby accelerates eddy motions. In addition, the 
accelerated eddies induce stronger overshooting that expands the convective layer, 
consistent with the trend seen in Fig.~\ref{LateralVmap}. Note also that this is consistent 
with what we find in the resolution dependence of PNS convection for $19\, M_{\sun}$ models; 
the low-resolution model experiences a bit stronger PNS convection than the others. The failure 
of the shock to explode in the low-resolution model results in greater ongoing mass accretion 
onto the PNS core, and, thus, a slightly more massive PNS. The correlation of PNS mass and average accretion rate with the vigor 
of PNS convection is a useful means with which to analyze the progenitor dependence of PNS convection.

\subsection{Turbulence in PNS convection}\label{subsec:turbulence}

\begin{figure*}
  \rotatebox{0}{
    \begin{minipage}{0.9\hsize}
	\includegraphics[width=\columnwidth]{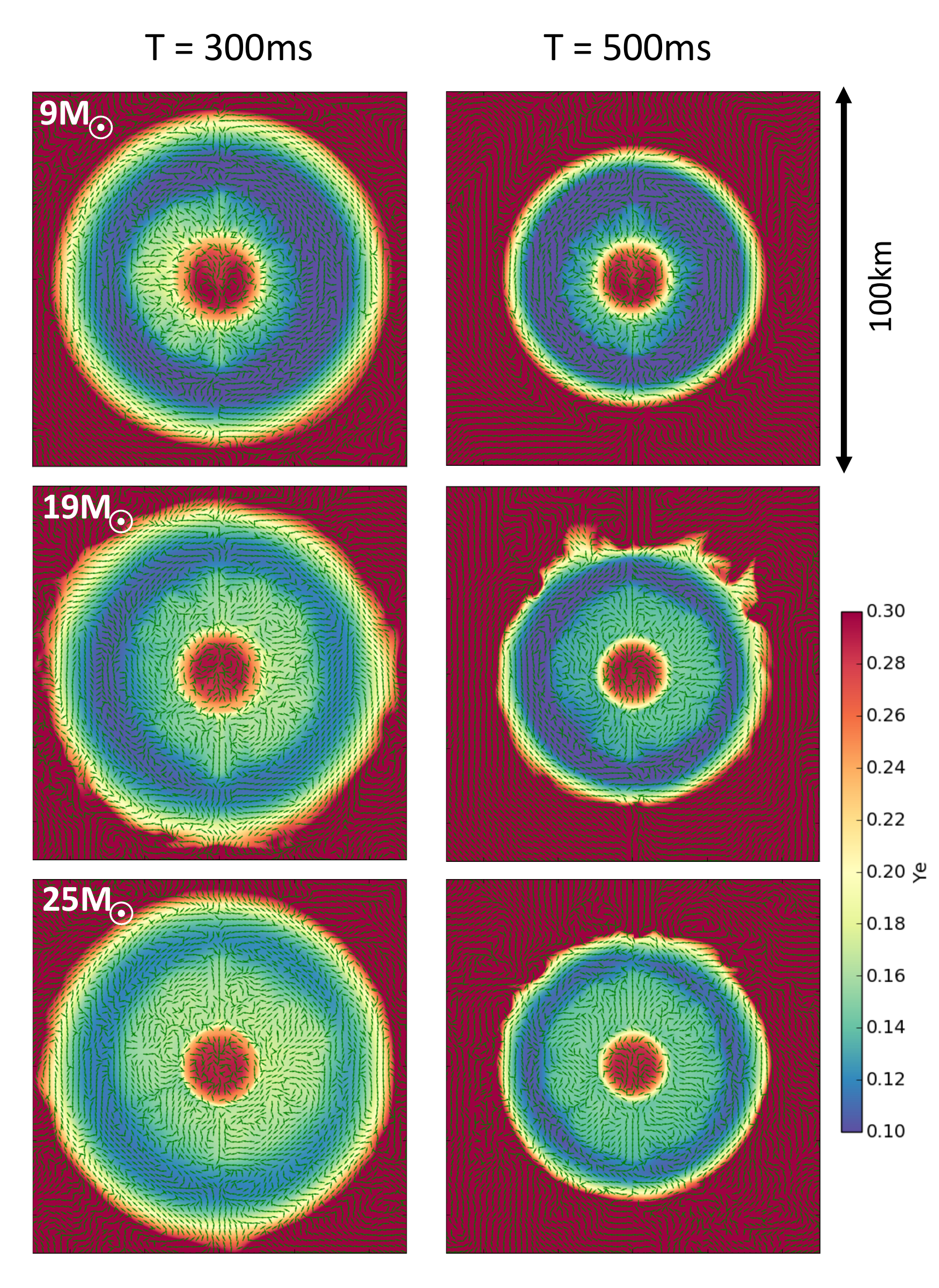}
    \caption{Two dimensional (X-Z) slice map of $Y_e$ in color, with fluid motions as vectors. 
Note that the length of each velocity vector is forced to be the same. The models
and times shown are the same as in Fig.~\ref{graph_3Dvisualization}.}
    \label{graph_2D_xz_visualization}
  \end{minipage}}
\end{figure*}

\begin{figure}
  \begin{minipage}{1.0\hsize}
    \includegraphics[width=\columnwidth]{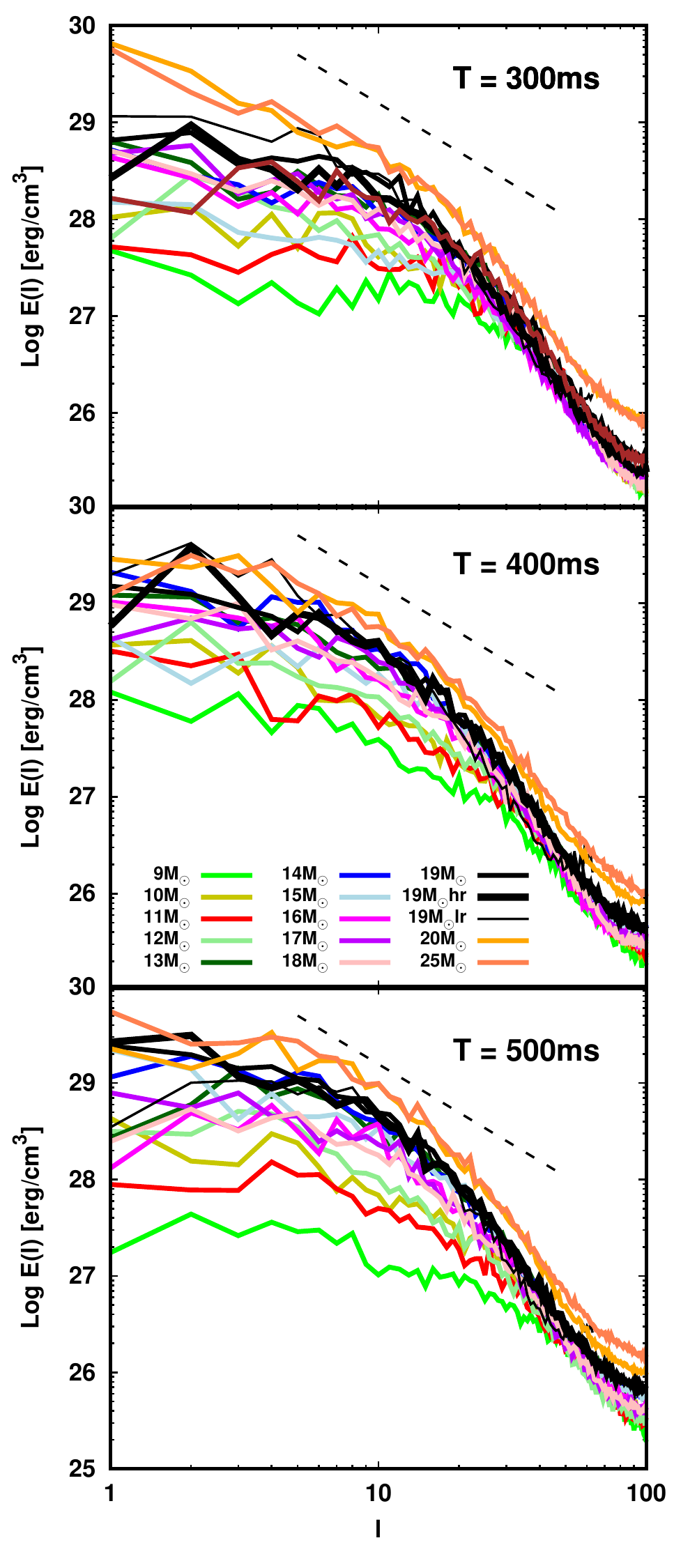}
    \caption{The power spectrum of the transverse kinetic energy density versus spherical harmonic 
index ($\ell$). It is computed and averaged in the region between two isodensity surfaces, $10^{13}$ 
and $10^{14}\, {\rm g/cm^3}$, which roughly corresponds to the PNS convective layer (Fig.~\ref{LateralVmap}). 
The power spectrum is averaged in time over an interval of $10$ ms. Note that we do not include the result for the 
$60\, M_{\sun}$ model, except for the time snapshot at $T=300$ ms after bounce (top panel), since that simulation was terminated 
before $T=400$ ms. The dashed line has a slope of $-\frac{5}{3}$ so as to compare with a Kolmogorov spectrum.}
    \label{graph_TurbulentSpectrum_10msave_Pdepe_in}
  \end{minipage}
\end{figure}

\begin{figure*}
  \rotatebox{90}{
    \begin{minipage}{1.3\hsize}
	\includegraphics[width=\columnwidth]{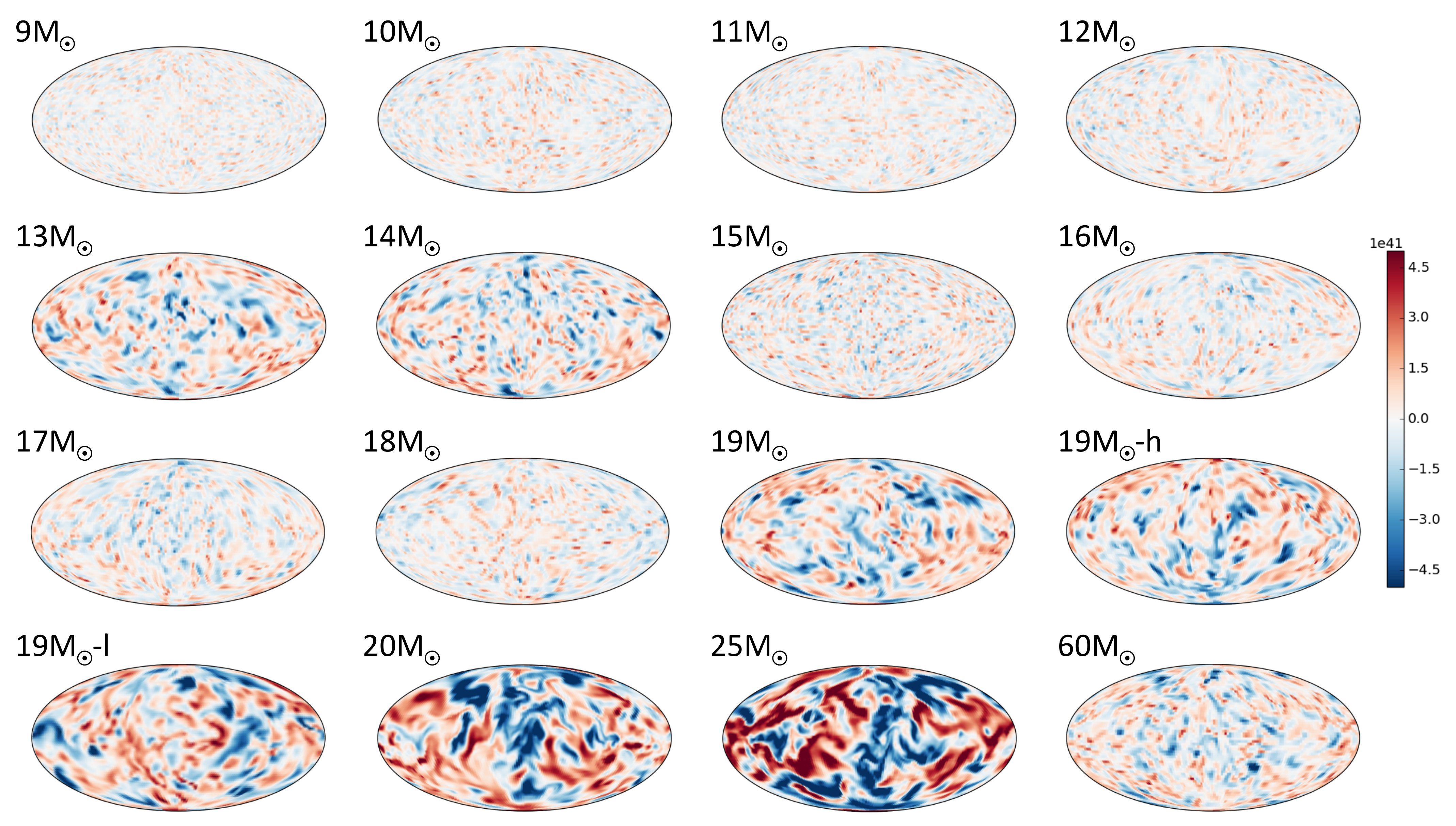}
    \caption{Mollweide projections of the total turbulent energy flux at a radius where $\rho_{\rm ave} = 5 \times 10^{13} {\rm g/cm^3}$. 
This position roughly coincides with the center of PNS convection at $300$ ms after bounce. The units are [${\rm erg/cm^2 s}$]. 
Note that the greater color contrast indicates a higher variation in the turbulent flux.}
    \label{MollweidePro_rho5e13}
  \end{minipage}}
\end{figure*}

\begin{figure}
  \begin{minipage}{1.0\hsize}
    \includegraphics[width=\columnwidth]{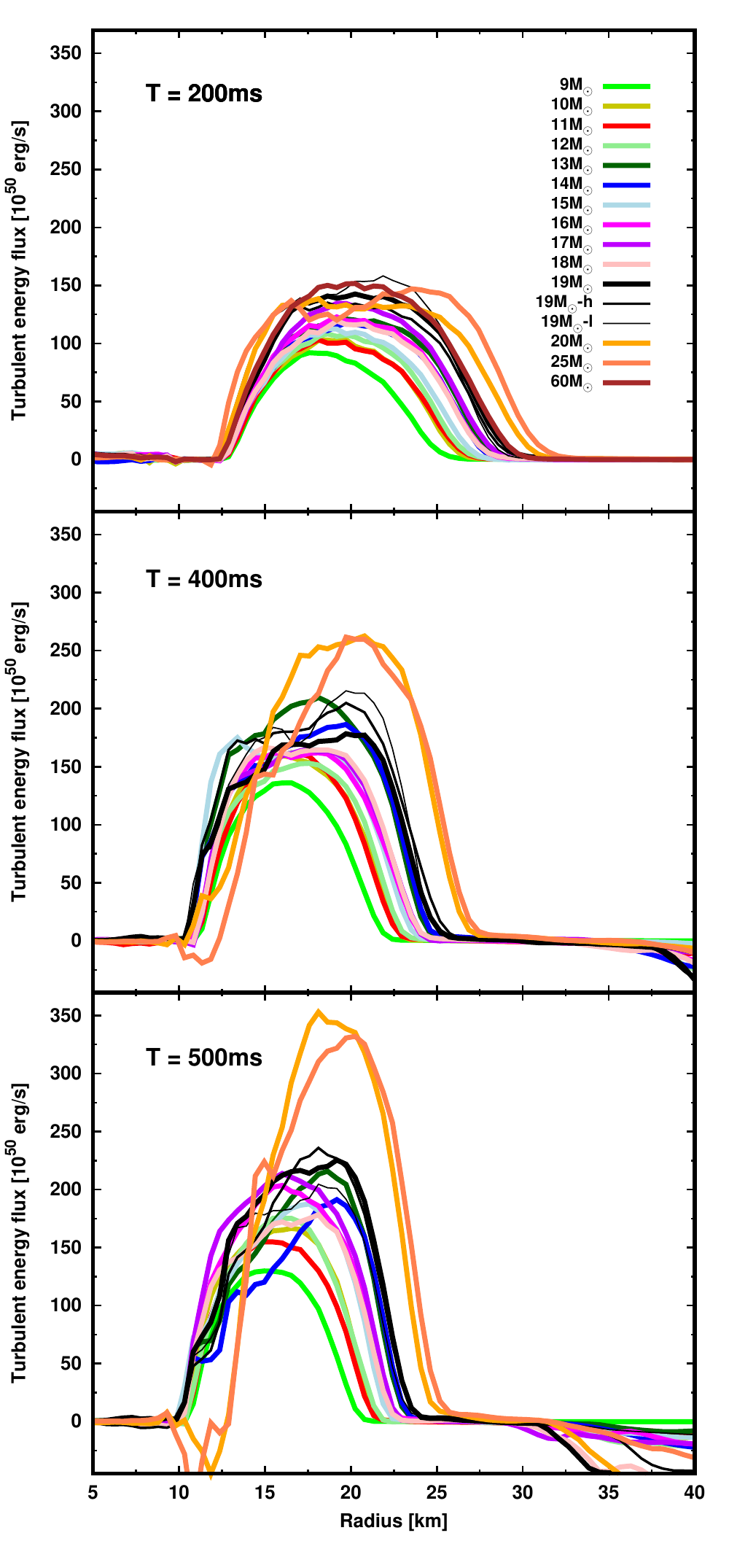}
    \caption{The radial profile of the angle-averaged turbulent energy flux at $200, 400$ and $500$ ms after 
bounce from top to bottom, respectively. The results for the $60 M_{\sun}$ model are not included in the last two snapshots.}
    \label{graph_TurbulentFlux}
  \end{minipage}
\end{figure}

\begin{figure}
  \begin{minipage}{1.0\hsize}
    \includegraphics[width=\columnwidth]{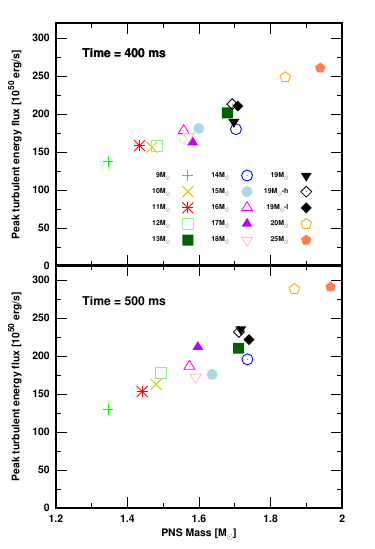}
    \caption{The same as Fig.~\ref{graph_PNSmass_vs_kineticEne}, but for the peak turbulent energy flux.}
    \label{graph_PNSmass_vs_Turbfluxmax}
  \end{minipage}
\end{figure}

We observe that PNS convection starts near $\lesssim 50$ ms after bounce and is sustained for the 
rest of the post-bounce phase. Convective motions grow rapidly, quickly transitioning to the non-linear
turbulent phase. However, the spectrum of turbulence in PNS convection is
not described by the simple Kolmogorov power-law. We here characterize the turbulent flow in 
PNS convection and then discuss its progenitor dependence.

Figure~\ref{graph_2D_xz_visualization} displays X-Z plane slice colormaps of the $Y_e$ distributions,
superposing fluid velocity vector fields. We observe that the inhomogeneities in $Y_e$ coincide with  
the eddy structure traced by the vector fields. In the linear phase of 
PNS convection, high-$Y_e$ bubbles move upward from the surface of the PNS core ($\sim$10 km)
and low-$Y_e$ structures sink down from the outer edge of the convective layer. Both flows reflect at the 
convective boundaries and execute circular motions. This results in a maximum curvature radius for 
the eddies equal to roughly the half width of the convective layer. Once the convection becomes non-linear,
smaller eddies emerge through the collision of larger eddies. Such dynamics is consistent with the 
classical process of the turbulent cascade.

As an aside, we remark on what Figure~\ref{graph_2D_xz_visualization} suggests about the LESA. 
As shown, a large-scale (dipole) asymmetry in the $Y_e$ distribution is clearly seen
in the $9-$ and $19-M_{\sun}$ models, associated with what we observed earlier in our published models 
\citep{2019MNRAS.489.2227V}. However, contrary to the previous claim concerning the possible physical origin
of the LESA phenomenon by \citet{2019ApJ...881...36G}, we do not observe concomitant large-scale 
coherent matter motions in the envelope of the PNS. It should be noted that 
large-scale coherent motions can be triggered by the random breaking of global 
symmetry in the shocked envelope, not necessarily first in the PNS convective zone, and 
that asymmetries in the latter might be a consequence of the former. Moreover, such 
symmetry breaking might be correlated with the PNS kick direction, as suggested by \citet{2019ApJ...880L..28N}.
Indeed, PNS proper motion should be a generic outcome of a broken symmetry 
which might be connected to the LESA direction and/or the
vector dipole direction of the exploding shock \citep{2019MNRAS.489.2227V}.
However, we suggest that a comprehensive understanding of the origin of the LESA,
its direction, and potential correlations with the vector directions of other phenomena
has yet to be found and we will leave this topic to future work.

To investigate the turbulent state of PNS convection, we compute the turbulent kinetic energy 
spectrum as a function of spherical-harmonics index $\ell$ at three different times; the 
results are displayed in Fig.~\ref{graph_TurbulentSpectrum_10msave_Pdepe_in}. We use the 
formalism of \citet{2015ApJ...808...70A}, but the region sampled and averaged is between two isodensity 
surfaces: $10^{13}$ and $10^{14}\, {\rm g/cm^3}$. As displayed in this figure, the spectra obey a broken power-law, 
which indicates that smaller eddies are actually created in the non-linear phase. It should be noted, however, 
that the spectra deviate from Kolmogorov's theory. Indeed, they decline more sharply than $\ell^{-\frac{3}{5}}$ at $\ell > 10$. 
This behavior seems not due to numerical resolution, since there are no remarkable differences in 
the spectra for the three different resolution models of the $19\, M_{\sun}$ progenitor \citep{2019MNRAS.tmp.2343N}. In 
addition, we find that the spectrum at lower $\ell$s ($\ell \lesssim 10$) strongly varies with progenitor. Models with 
stronger PNS convection tend to evince a sharper decline in the spectrum $-$ turbulent kinetic energy accumulates in the 
large eddies. Some possible explanation for this tendency might be found by observing the systematics
of PNS convection with radial width. As mentioned already, models with stronger PNS convection tend 
to have convective layers with wider radial widths (Fig.~\ref{LateralVmap}), which results in expanding the 
maximum eddy scale in the convective region. As a result, stronger convection generates 
larger eddies more efficiently. The deviation from the Kolmogorov's cascade at higher $\ell$(>10) 
may also imply that momentum transfer by neutrinos, or neutrino-drag, affects the spectrum. 
Indeed, neutrino diffusion is more efficient for smaller eddies, i.e., the effects of neutrino 
viscosity are greater for smaller eddies 
\citep{1984A&A...136...74V,1988PhR...163...51B,1993ApJ...408..194T,2015MNRAS.447.3992G,2019arXiv190401699M}. 
We also note that the small to modest aspect ratio of the convective region may result in a quasi-2D
cascade, i.e. inverse from small to large scales.  However, we don't yet feel comfortable, in lieu of even higher
resolution studies, highlighting this possibility at this time.

In Fig.~\ref{MollweidePro_rho5e13}, we provide angular maps of the turbulent energy 
flux (kinetic and thermal energy flux) in the PNS zone. Note that the total energy 
flux is dominated by the thermal contribution, This is consistent with the fact that
the turbulent Mach number in PNS convection is less than $10 \%$, i.e., the kinetic 
energy is less than a few $\%$ (\S\ref{subsec:basicchara}). In Fig.~\ref{MollweidePro_rho5e13}, 
a Mollweide projection of the turbulent energy flux at $300$ ms after core bounce is 
displayed. This is computed at the isodensity surface where the matter density 
is $5 \times 10^{13} {\rm g/cm^3}$, which is roughly located in the middle of the 
convective layer. As shown in the figure, when overall
PNS convection is stronger, there are higher angular variations. Indeed, the $25\, M_{\sun}$ model 
has the strongest color contrast of all the maps, while the $9\, M_{\sun}$ model has the weakest contrast. 
These figures also clearly indicate that models with stronger PNS convection have larger scale  
variations in the angular directions. This is consistent with our finding that the turbulent-kinetic-energy 
spectra at lower $\ell$s possess significantly higher energies than at higher $\ell$s
(Fig.~\ref{graph_TurbulentSpectrum_10msave_Pdepe_in}).

Figure ~\ref{graph_TurbulentFlux} of the angle-averaged turbulent energy flux 
at three different time snapshots clearly shows that PNS convection commonly generates 
net outgoing energy fluxes. This is not a trivial outcome and could
be of relevance in the possible generation of hydrodynamic waves \citep{2019arXiv191007599G}.  
Indeed, as Fig.~\ref{MollweidePro_rho5e13} 
demonstrates, incoming energy flux dominates outgoing flux at some angular points. The 
dominance of the outgoing energy flux indicates that energy transport by buoyancy 
is more efficient than that by downflows from the outer edge of the convective layer. This 
is in part due to the inhomogeneous matter distributions in background flows, when there is a 
strong negative density gradient in the convective layer. The thermal enthalpy is 
also higher at smaller radii, indicating that energy transport by buoyant motions 
is larger than that of downflows. It is also important to mention that the 
energy flux is time-dependent and that the peak flux grows with time for the duration of our 
simulations. In accordance with the PNS contraction, the radial width of the outgoing flux 
also changes. We note that in the turbulent gain region before explosion, the corresponding
net turbulent energy flux is negative, not positive.  This significant difference 
is due to the modest-to-large post-shock accretion/infall speeds behind the shock 
as the accreted matter settles onto the inner core. This difference highlights 
one important distinct characteristic of PNS convection $-$ it is in a truly ``stellar"
convective zone in the classic stellar evolution sense.

Figure~\ref{graph_TurbulentFlux} also suggests that models with stronger PNS                                                                         
convection manifest higher angle-averaged turbulent energy flux. Indeed, the $25\, M_{\sun}$ model                                                        
has the highest peak turbulent energy flux. To further buttress our argument that the                                                                 
PNS mass and the vigor of PNS convection are correlated, we compare the PNS mass and                                                                      
the peak angle-averaged turbulent energy flux in Fig.~\ref{graph_PNSmass_vs_Turbfluxmax}.                                                                 
This figure clearly displays this positive correlation and lends confidence to our finding                                                                
that the PNS mass is an important determinant of the vigor of PNS convection.

\subsection{Neutrino emissions}\label{subsec:neutrinoemis}

\begin{figure*}
  \begin{minipage}{0.9\hsize}
    \includegraphics[width=\columnwidth]{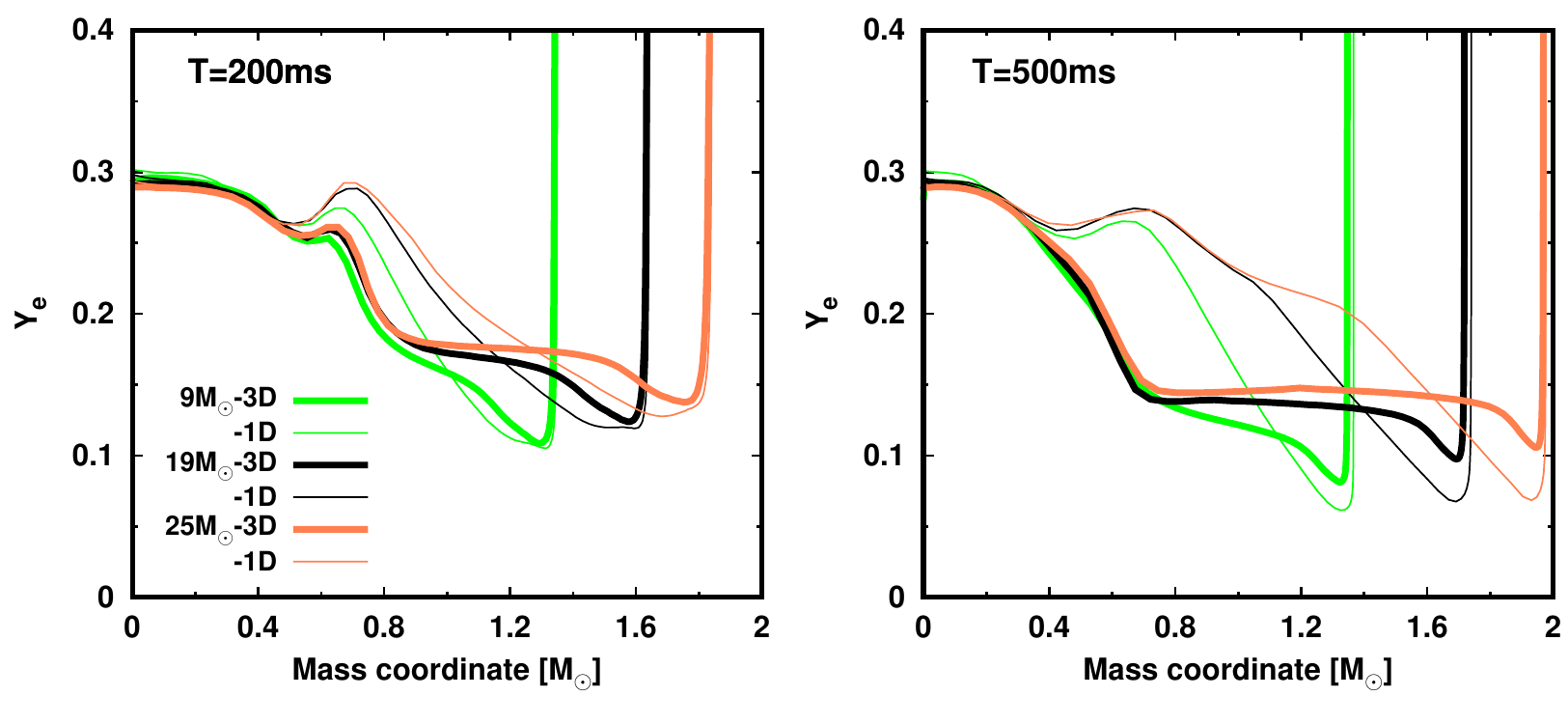}
    \caption{Angle-averaged $Y_e$ profile as a function of enclosed mass for 
representative 3D models (thick lines). The corresponding profiles obtained from 
spherically-symmetric simulations (thin lines) are also displayed.}
    \label{graph_Ye_profile_asM_t200ms500ms}
  \end{minipage}
\end{figure*}

\begin{figure}
  \begin{minipage}{1.0\hsize}
    \includegraphics[width=\columnwidth]{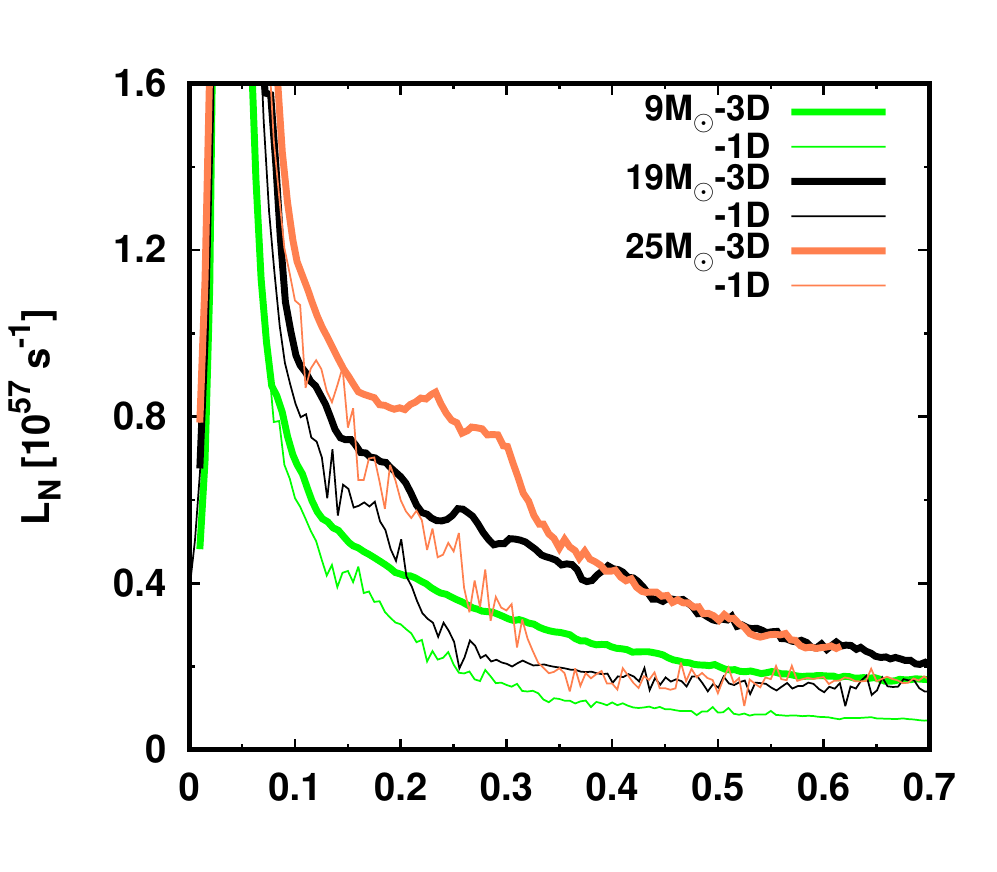}
    \caption{The time evolution of electron-lepton number luminosity for the three selected models.}
    \label{graph_neutrino_ELNflux}
  \end{minipage}
\end{figure}

\begin{figure*}
  \begin{minipage}{0.9\hsize}
    \includegraphics[width=\columnwidth]{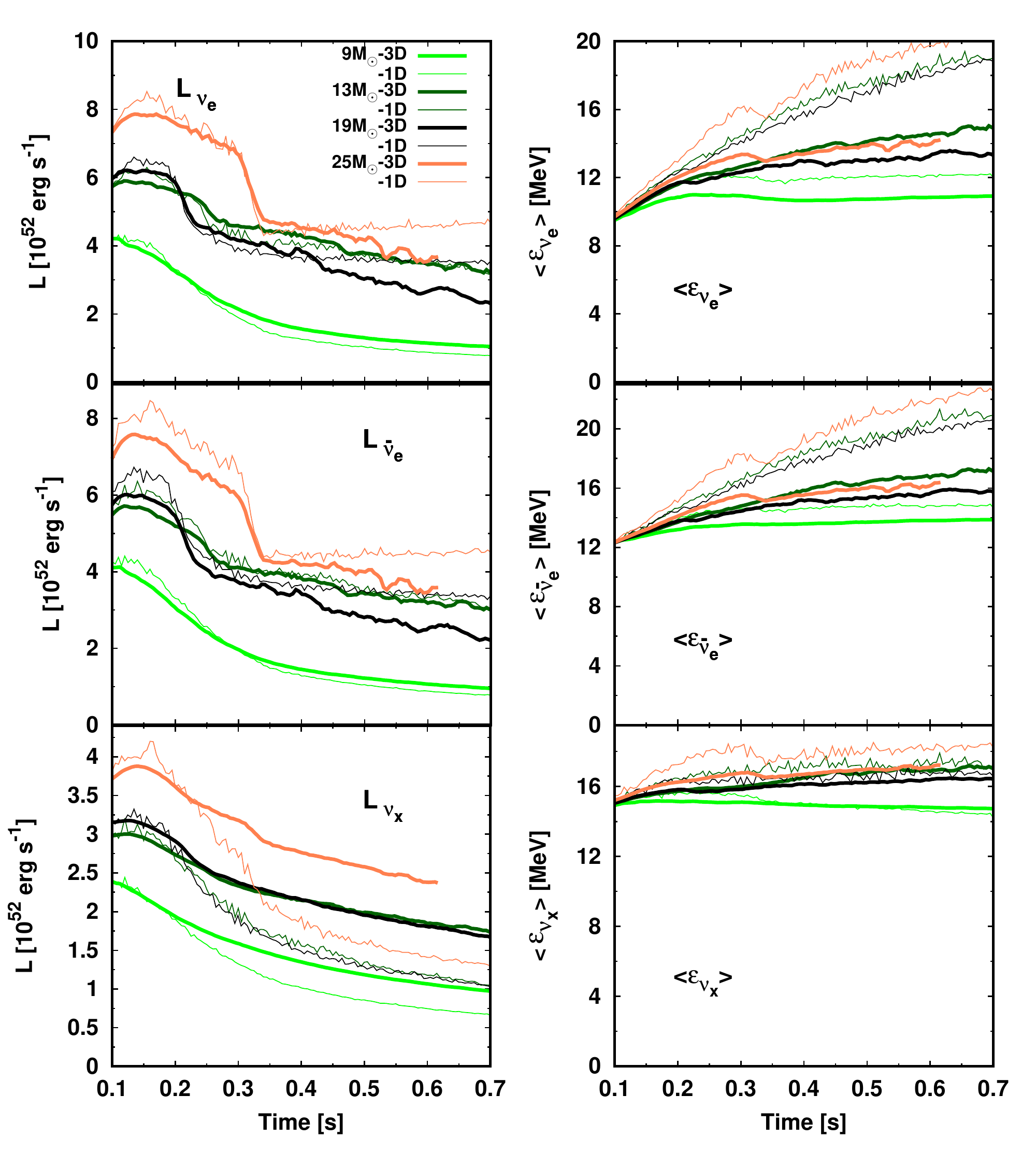}
    \caption{The energy luminosity (left) and average neutrino energy (right) as a function of time for each 
species of neutrino for four selected models. Note that the $13 M_{\sun}$ model 
represents models that didn't explode in this suite of 3D simulations. The values are taken at $10,000$ km.}
    \label{graph_neutrino_lumi_aveene_PNS}
  \end{minipage}
\end{figure*}


The effect of PNS convection on neutrino emissions is one of the fundamental issues in CCSN theory.
We focus in our analysis on a subset of representative progenitors and compare our 3D results with the results 
for corresponding spherically-symmetric simulations to quantify the impact of PNS convection.

First, we consider the effect of PNS convection on deleptonization. Fig.~\ref{graph_Ye_profile_asM_t200ms500ms} 
displays radial profiles of angle-average $Y_e$ distributions for three selected progenitors ($9-, 19-$, 
and $25-M_{\sun}$) as a function of interior baryon mass coordinate\footnote{See also the top panels of 
Fig.~\ref{graph_Ye_S_profile_asR_t200ms500ms}, which display the same quantities but as a function of radius.}. 
As shown in the figure, $Y_e$ at $M \gtrsim 0.4\, M_{\sun}$ is quite different 
from that for the spherically-symmetric simulations. In the vicinity of the inner edge of the convective 
layer $Y_e$ is systematically smaller in 3D than in 1D, whereas there is the opposite behavior at the outer edge. 
This is evidence that PNS convection carries the leptons (mainly electrons) from the inside to the outside of 
the convective zone. In addition, we find that the total electron number inside the PNS is smaller in 
3D than in 1D\footnote{The electron number can be computed using $\int dM Y_e(M)$, which corresponds to the 
area in the $Y_e-M$ figure.}, i.e., PNS convection works to accelerate deleptonization of the PNS. More 
quantitatively, the core deleptonization rate is roughly two times higher in 3D than in 1D. This can be seen 
using the electron lepton number (ELN) luminosity displayed in Fig.~\ref{graph_neutrino_ELNflux}. 
This phenomenon is mainly due to the fact that PNS convection carries electrons (and $\nu_e$s) from very optically 
thick regions to the relatively thin regions. We note that at the outer edge of the PNS convective zone 
most $\nu_e$s are still trapped, but that the lower-energy neutrinos preferentially escape,  
enhancing deleptonization. We find, however, that the difference in the ELN luminosity 
between 1D and 3D for the $19-$ and $25-M_{\sun}$ models moderates with time.  This is in part 
due to the fact that $\nu_e$ and $\bar{\nu}_e$ emissions undergo a sharp drop at shock 
revival. Although there is some progenitor dependence of core deleptonization, 
we confirm that in general PNS convection accelerates it. This is consistent with 
the results of previous 2D simulations \citep{2006A&A...457..281B}.

Figure~\ref{graph_neutrino_lumi_aveene_PNS} displays the time evolution of the neutrino luminosity and 
the average neutrino energies for the $9-, 13-, 19-$ and $25-M_{\sun}$ progenitors for both 3D and 1D models; 
the $13\, M_{\sun}$ model is added to this figure to represent non-exploding models. In the early 
post-bounce phase ($\lesssim 200$ ms), we find that the 3D neutrino luminosities for all species 
are slightly smaller than in 1D, with the difference being smaller still for the $9\, M_{\sun}$ model.
Later, the difference in neutrino luminosity between 1D and 3D depends upon 
model and species. We now describe the origin of this complexity.

As shown in Figs.~\ref{graph_Ye_S_profile_asR_t200ms500ms}~and~\ref{graph_Ye_profile_asM_t200ms500ms}, 
the matter distributions in the envelope of the PNS in 3D are already different from those 
in 1D, even during the early post-bounce phase ($\lesssim 200$ ms). This affects the locations of the various 
neutrinospheres and the thermodynamic state of the matter there. As displayed in 
Fig.~\ref{graph_PNSradius_1Dvs3D}, the PNS radius in 3D models is systematically larger 
than in 1D. This would work to increase the neutrino luminosity, whereas the matter temperature 
at the neutrinospheres is smaller than in 1D, working to decrease
the luminosity\footnote{Consistent with the fact that the matter temperatures in 3D are lower here 
than in 1D is that the average energy of the neutrinos in 3D is smaller than in 1D 
(Fig.~\ref{graph_neutrino_lumi_aveene_PNS}).}. This competition can be characterized very roughly
by a parameter, $\beta$, which is defined as 
\begin{eqnarray}
\beta \equiv \frac{L_{\nu 3D}}{L_{\nu 1D}} \sim \frac{{\rm T}^{4}_{\nu 3D} {\rm R}^{2}_{\nu 3D}}{{\rm T}^{4}_{\nu 1D} {\rm R}^{2}_{\nu 1D}}\, ,
\label{eq:Lcompete}
\end{eqnarray}
where ${\rm T}_{\nu}$ and ${\rm R}_{\nu}$ denote the matter temperature and radius, respectively, of the 
neutrinosphere\footnote{Note that the effect of Fermi degeneracy is not taken into account in Eq.~\ref{eq:Lcompete}
(see e.g., \citet{2013ApJ...765..123N}).}. We find that during the early phase the latter effect dominates the former ($\beta < 1$). 
As a result, during this early phase the neutrino luminosity is smaller in 3D. Unlike the case for the other 
species, the production of $\bar{\nu}_e$s is suppressed by the Fermi degeneracy of the $\nu_e$s.
Due to the enhanced supply of leptons by PNS convection, this suppressive effect
on the density of $\bar{\nu}_e$s is stronger in 3D than in 1D \citep{2006A&A...457..281B}.

At the later phase ($\gtrsim 400$ ms), the difference in neutrino luminosity between 1D and 3D 
has a different origin than during the earlier phase. The $\nu_x$ luminosity in 3D is higher 
than in 1D for all models, but the differences between 1D and 3D models for the $\nu_e$ and $\bar{\nu}_e$ 
neutrinos depend upon progenitor. The luminosities of the latter species are lower in 3D than 1D for the $19-$ and $25-M_{\sun}$ 
models, but higher in $9\, M_{\sun}$ model. On the other hand, the $\nu_e$ and $\bar{\nu}_e$ 
luminosities for the $13-M_{\sun}$ model depend little on dimension. Such a complex 
progenitor dependence is due to competing effects.

Unlike during the early post-bounce phase, the effect on the luminosity of the physical 
expansion of each neutrinosphere as a result of PNS convection dominates that due to the 
countervailing effect of the corresponding reduction of the matter temperature, i.e., $\beta$ tends 
to be greater than unity in the later phase. However, the effect of Fermi degeneracy on 
the $\bar{\nu}_e$ luminosity mutes the increase in luminosity. Therefore, the net effect of PNS convection 
on the $\bar{\nu}_e$ luminosity is smaller than that for the other species. This makes subtle the difference between 
1D and 3D for the $\bar{\nu}_e$ luminosity and accounts for what we witness for the $9 M_{\sun}$ 
model. Since the accretion component of the neutrino luminosity is negligible at later times 
due to the swift onset of a vigorous explosion, this conclusion for the $9 M_{\sun}$ model 
is that much more robust.  However, the other, more massive, progenitors have higher mass 
accretion rates (Fig.~\ref{graph_Massaccretion}). Thus, effects of the accretion component 
on the neutrino luminosities is for them of more import. For explosion 
models (e.g., the $19-$ and $25-M_{\sun}$ models), the mass accretion rate is reduced by the revival of the shock 
(Fig.~\ref{graph_Massaccretion}). Hence, the accretion component of the neutrino luminosity is reduced 
substantially. For the $\nu_e$ and $\bar{\nu}_e$ neutrinos, this effect overwhelms the increase 
of the luminosity by PNS convection, which results in a lower luminosity in 3D than in 1D. This conclusion 
is buttressed by the results for the non-exploding models (e.g., the $13-, 14-, {\rm{and}}15-M_{\sun}$ models), in which the 
difference between 1D and 3D in both the $\nu_e$ and $\bar{\nu}_e$ luminosities is smaller than that in 
the exploding models. On the other hand, the $\nu_x$ neutrino luminosity is always higher in 3D than in 1D.
This is in part attributed to the fact that the accretion component of the $\nu_x$ luminosity is smaller 
than the diffusion component, regardless of progenitor.

Finally, we discuss the impact of PNS convection on CCSN dynamics. As 
mentioned above, PNS convection does not much change the luminosity of the $\nu_e$ and $\bar{\nu}_e$ 
neutrinos, but has an influence on their average energies; these are reduced more than $10 \%$. The 
reduction of these average energies suggests that the efficiency of neutrino heating in the gain region 
might be reduced, i.e., PNS convection might negatively affect shock revival.  However, the 
enhanced $\nu_x$ luminosity accelerates the shrinkage of the PNS itself 
\citep{2017ApJ...850...43R} and this effect, like the effect of general relativity \citep{2004ApJS..150..263L} 
and the many-body corrections to the neutrino-matter interaction rates \citep{2016arXiv161105859B}, aids explosion.  
Hence, since there are numerous countervailing effects of PNS convection on core evolution 
and the neutrino emissions, the actual net impact of PNS convection on the explosion for all 
progenitor structures is still unclear; a more detailed study is called for to resolve all the 
effects, and the net effect, of PNS convection on CCSN dynamics.

\section{Summary and Discussion}\label{sec:sumanddiscuss}

In this paper, we have carried out the first systematic study of PNS convection in 3D for a broad
and representative range of progenitor masses. We confirm that PNS convection is a generic, persistent, 
and important feature of all models and that there is a rich diversity among progenitors. We 
find in the first $\sim$second after bounce that the overall dynamics and behavior of PNS 
convection is insensitive to whether the shock is revived. Moreover, we find that the 
mass of the PNS is the most important quantity that characterizes the vigor of convection 
and that more massive PNS cores experience more vigorous convection. Moreover, stronger PNS convection 
results in a physically larger convective layer. We also find that PNS convection in 
simulations with different resolutions has similar characteristics.
We speculate that at later times ($> 1$s after core bounce) PNS convection 
will manifestly differ between models that fail to explode and models that don't, and may be 
very interestingly different near black hole formation. Longer simulations are definitely 
required to assess such differences. 

We analyzed the characteristics of the turbulence in PNS convection and found that in 3D it
deviated from Kolmogorov's theory. Also, we found that the turbulence is anisotropic; indeed the net 
turbulent energy flux is outgoing and we confirm that PNS convection 
accelerates the deleptonization of the PNS. The ELN luminosity of the core in 3D is roughly two times higher than 
that in 1D models (Fig.~\ref{graph_neutrino_ELNflux}). This is attributed to the fact that 
PNS convection dredges up leptons to the boundary of the PNS convective zone and, thereby enhances the lepton flux
at the outer edge of the PNS. Furthermore, the expanded neutrinospheric radii, accompanied by lower 
neutrinospheric temperatures compared with those in 1D, accelerates total ELN loss from the PNS. As a result, there are 
important differences in neutrino emissions between 1D and 3D that also depend strongly
on progenitor. Complexity arises, however, from the competition of factors, which we have attempted to clarify in this 
study. 

Although the results presented in this paper are based on a state-of-the-art methodology,
uncertainties in some of the input physics, in particular, some neutrino opacities \citep{2016arXiv161105859B} 
and the nuclear EOS \citep{2012PhRvL.108f1103R,2013ApJ...774...17S,2017RvMP...89a5007O}, may affect 
our findings. As we have observed, PNS convection occurs in the density 
range of $10^{13} \lesssim \rho \lesssim 10^{14}\, {\rm g/cm^3}$, which corresponds to the transition 
region from inhomogeneous to uniform hadronic matter. In this region, there still remain many 
ambiguities in both the EOS and the weak-interaction rates. We have not yet determined the sensitivity of PNS 
convection to these inputs, and doing so should be a goal of future research.

We note that the convective zone in the PNS is an ideal context for the dynamo amplification
of magnetic fields, similar to what happens in convective stars and planets.  PNS convection persists
independent of explosion details and an $\alpha^2$ and/or $\alpha-\Omega$ (if rotating)
dynamo would seem naturally to arise in this generic convective context.  Equating 1\% of the kinetic energies
in the convective layers that we find in our 3D simulations (Fig.~\ref{graph_tevo_PNSconv_kineticEne})
to a mean magnetic energy density times the calculated volume of the convective zones reveals that
an ``equipartition" magnetic field of 10$^{12}$ to 10$^{15}$ gauss can easily be accommodated.
It can even be larger, but since this region is progressively buried and has a mass cap
over it when the neutron star effectively ceases accretion \citep{2019arXiv190904152B} we might expect
the final surface fields to be smaller.  Furthermore, generating B-fields in a shell should
favor higher-order harmonics locally and the dipole fraction, the one most relevant 
to pulsar astronomy, is likely subdominant at depth. Therefore, what the surface 
dipolar moment might be would be an important issue. Finally, 
the systematics in the vigor and the extent of PNS convection with progenitor structure 
and PNS mass may point to a corresponding systematics in the B-fields of pulsars born 
from them.  This intriguing possibility, that a pulsar's magnetic fields originates 
in PNS convection, deserves further scrutiny, but shows promise.

\section*{Acknowledgements}

The authors acknowledge ongoing contributions to
this effort by Josh Dolence and Aaron Skinner.
We also acknowledge Evan O'Connor regarding the equation of state,
Gabriel Mart\'inez-Pinedo concerning electron capture on heavy nuclei,
Tug Sukhbold and Stan Woosley for providing details concerning the
initial models, Todd Thompson regarding inelastic scattering,
and Andrina Nicola for help in computing the turbulent spectrum.
We acknowledge support
from the U.S. Department of Energy Office of Science and the Office
of Advanced Scientific Computing Research via the Scientific Discovery
through Advanced Computing (SciDAC4) program and Grant DE-SC0018297
(subaward 00009650). In addition, we gratefully acknowledge support
from the U.S. NSF under Grants AST-1714267 and PHY-1144374 (the latter
via the Max-Planck/Princeton Center (MPPC) for Plasma Physics). DR
cites partial support as a Frank and Peggy Taplin Fellow at
the Institute for Advanced Study. An award of computer time was provided 
by the INCITE program. That research used resources of the
Argonne Leadership Computing Facility, which is a DOE Office of Science 
User Facility supported under Contract DE-AC02-06CH11357. In addition, this overall research 
project is part of the Blue Waters sustained-petascale computing project,
which is supported by the National Science Foundation (awards OCI-0725070
and ACI-1238993) and the state of Illinois. Blue Waters is a joint effort
of the University of Illinois at Urbana-Champaign and its National Center
for Supercomputing Applications. This general project is also part of
the ``Three-Dimensional Simulations of Core-Collapse Supernovae" PRAC
allocation support by the National Science Foundation (under award \#OAC-1809073).
Moreover, access under the local award \#TG-AST170045
to the resource Stampede2 in the Extreme Science and Engineering Discovery
Environment (XSEDE), which is supported by National Science Foundation grant
number ACI-1548562, was crucial to the completion of this work.  Finally,
the authors employed computational resources provided by the TIGRESS high
performance computer center at Princeton University, which is jointly
supported by the Princeton Institute for Computational Science and
Engineering (PICSciE) and the Princeton University Office of Information
Technology, and acknowledge our continuing allocation at the National
Energy Research Scientific Computing Center (NERSC), which is
supported by the Office of Science of the US Department of Energy
(DOE) under contract DE-AC03-76SF00098.





\bibliographystyle{mnras}
\bibliography{bibfile}







\bsp	
\label{lastpage}
\end{document}